\documentclass[12pt]{article}
\usepackage[utf8]{inputenc}

\usepackage{adjustbox}
\usepackage{authblk}
\usepackage[utf8]{inputenc}
\usepackage{xcolor}
\usepackage{graphicx}
\usepackage{bm}
\usepackage{amsmath}

\numberwithin{equation}{section}
\usepackage{pbox}
\usepackage{physics}
\usepackage{amssymb }
\usepackage{array}
\usepackage{tikz}
\usetikzlibrary{shapes.geometric}
\usetikzlibrary{calc,patterns,angles,quotes}
\usetikzlibrary{decorations.pathreplacing}
\usepackage{hyperref}
\hypersetup{
    colorlinks,
    citecolor=blue,
    filecolor=black,
    linkcolor=black,
    urlcolor=blue
}
\hypersetup{linktocpage}
\usepackage{caption}
\usepackage[mathscr]{euscript}
\usepackage[backend=biber,style=numeric-comp,sorting=none,maxbibnames=99]{biblatex}
\usepackage{scalerel}
\usepackage{stackengine,wasysym}
\usepackage[margin=1in]{geometry}

\def\la{\lambda}

\def\si{\sigma}
\def\ga{\gamma}
\def\sp{\hspace{0.05cm}}
\def\spa{\hspace{0.1cm}}
\def\spac{\hspace{1cm}}
\newcommand\reallywidetilde[1]{\ThisStyle{%
  \setbox0=\hbox{$\SavedStyle#1$}%
  \stackengine{-.1\LMpt}{$\SavedStyle#1$}{%
    \stretchto{\scaleto{\SavedStyle\mkern.2mu\AC}{.5150\wd0}}{.6\ht0}%
  }{O}{c}{F}{T}{S}%
}}

\begin{document}
\baselineskip 5mm

\begin{titlepage}

  \hfill
  \pbox{10cm}{
    \texttt{HU-Mathematik-2020-03}\\
    \texttt{HU-EP-20/13}\\
    \texttt{SAGEX-20-18-E}
  }
  
  \vspace{2\baselineskip}

  \begin{center}

    \textbf{\LARGE \mathversion{bold}
      The Dual Conformal Box Integral in Minkowski Space\\
    }

    \vspace{2\baselineskip}

    Luke Corcoran$ ^{ a}$ and Matthias Staudacher$ ^{ a,b}$
 
    \vspace{2\baselineskip}

    $^{a}$\textit{
     Institut für Mathematik und Institut für Physik, Humboldt-Universität zu Berlin,\\
      IRIS-Adlershof, Zum Großen Windkanal 6, 12489 Berlin, Germany\\
      \vspace{0.5\baselineskip}
    }
        \vspace{2\baselineskip}

    $^{b}$\textit{ Department of Physics, Ewha Womans University, \\
DaeHyun 11-1, Seoul 120-750, S. Korea
      \vspace{0.5\baselineskip}
    }

    \vspace{2\baselineskip}
    
     \texttt{
      \{corcoran,staudacher\}@physik.hu-berlin.de
    }

    \vspace{2\baselineskip}
\end{center}

\thispagestyle{empty}
\begin{abstract}\noindent The dual conformal box integral in Minkowski space is not fully determined by the conformal invariants $z$ and $\bar{z}$. Depending on the kinematic region its value is on a `branch' of the Bloch-Wigner function which occurs in the Euclidean case. Dual special conformal transformations in Minkowski space can change the kinematic region in such a way that the value of the integral jumps to another branch of this function, encoding a subtle breaking of dual conformal invariance for the integral. We classify conformally equivalent configurations of four points in compactified Minkowski space. We show that starting with any configuration, one can reach up to four branches of the integral using dual special conformal transformations. We also show that most configurations with real $z$ and $\bar{z}$ can be conformally mapped to a configuration in the same kinematic region with two points at infinity, where the box integral can be calculated directly in Minkowski space using only the residue theorem.
\end{abstract}
\end{titlepage}
\pagenumbering{arabic}
\tableofcontents
\section{Introduction}\label{Introduction}
\begin{figure}[h!]
\begin{center}
\begin{adjustbox}{max totalsize={.34\textwidth}{.27\textheight},center}
\begin{tikzpicture}{scale=0.2} 
\Large
\draw[dotted] (-2,2)--(-3.2,3.2); 
\draw[dotted] (2,2)--(3.2,3.2); 
\draw[dotted] (2,-2)--(3.2,-3.2);
\draw[dotted] (-2,-2)--(-3.2,-3.2); 
  \draw[dotted] (-2,2)--(2,2);     
  \draw[dotted] (-2,2)--(-2,-2);
  
  \draw[dotted] (-2,-2)--(2,-2);
 
  \draw[dotted] (2,-2)--(2,2);
\draw[thick] (-2.8,0)--(2.8,0);
\draw[thick] (0,-2.8)--(0,2.8);
\filldraw (0,0) circle (3pt);
\filldraw (-2.8,0) circle (3pt);
\filldraw (0,2.8) circle (3pt);
\filldraw (0,-2.8) circle (3pt);
\filldraw (2.8,0) circle (3pt);
  \node at (-0.5,2.8){$x_1$};
    \node at (-0.5,-2.8){$x_3$};  \node at (-2.8,0.4){$x_4$};  \node at (2.8,0.4){$x_2$};

    \normalsize
\end{tikzpicture}
\end{adjustbox}
\end{center}
\caption{One-loop massless box diagram in dual momentum space. Internal points are integrated over. Dotted lines show the diagram in momentum space.}
    \label{figg16}
\end{figure}
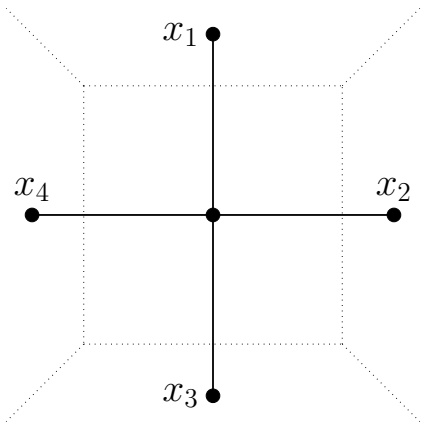

In recent years, there has been enormous progress with integrable conformal quantum field theories in dimensions higher than two. The most important example is planar $\mathcal{N}=4$ super-Yang-Mills theory (SYM) in four dimensions. Here an exact description exists at strong coupling due to the AdS/CFT correspondence \cite{Maldacena_1999,Gubser_1998,witten1998anti}. Moreover, the model appears to be integrable at arbitrary values of the coupling constant \cite{Beisert_2011}. While an exact proof is still missing, this apparent fact has been used to calculate a large number of quantities at weak, strong, and even finite coupling. What appears to be crucial for integrability is the conformal symmetry of the model in conjunction with a second, {\it dual} conformal symmetry \cite{Drummond_2007}. These two symmetries combine into an infinite-dimensional Yangian symmetry, which leads to integrability \cite{Ferro:2018ygf}. It should be stressed, from the derivation of the AdS/CFT correspondence, that the two copies of conformal symmetry should each be {\it global}, i.e.\ they should also be present for large, finite (ordinary and dual) conformal transformations in Minkowski space. A topical issue is to understand the integrability directly on the level of the Feynman diagrammatic expansion for correlation functions of $\mathcal{N}=4$ SYM. As we will review below, infinitesimal dual conformal symmetry indeed appears even on the level of the simplest radiative correction to four-point functions at one-loop order: The well-known box integral. However, vexingly, global invariance does not strictly hold for all large transformations. Although known, this rather subtle symmetry breaking has never been analysed in detail, and is the main focus of this paper: We relate the branched structure of the known analytic expression for the box integral to equivalence classes of four points in Minkowski-space with respect to conformal transformations at fixed values of the conformal cross-ratios. This required a systematic classification of the latter, which to the knowledge of the authors has not been worked out in full detail before. Furthermore, we studied a novel set of conformal four-point configurations in Minkowski space which cover all of the kinematic space. We describe the simplest of these configurations here, which involve moving two points (rather than the standard one point) to conformal infinity, and demonstrate how these configurations can simplify the calculation of the box integral. The consequences of our findings for the integrability program of $\mathcal{N}=4$ SYM will be studied in future work. 
\newline
\newline
The scalar one-loop massless box integral in dual momentum space (figure \ref{figg16}) is written
\begin{align}\label{2t}
    I(x_1,x_2,x_3,x_4)=\int \frac{d^4x_5}{i\pi^2}\frac{x_{13}^2x_{24}^2}{(x_{15}^2+i\epsilon)(x_{25}^2+i\epsilon)(x_{35}^2+i\epsilon)(x_{45}^2+i\epsilon)},
\end{align}
where $x_{ij}^2\equiv (x_i-x_j)^2$. $x_i\in \mathbb{R}^{1,3}$ are dual momenta \cite{Drummond_2010}, related to the usual QFT momenta\footnote{The $x_i$ can only be defined from $p_i$ up to an overall translation.} $p_i$ by $p_i=x_i-x_{i+1}$, with summation in the index taken mod 4. 
We take the $x_i$ to be sufficiently generic, so that in particular $x_{ij}^2\neq 0$ and we are away from light cone singularities\footnote{Away from these singularities the integral (\ref{2t}) is finite in four dimensions.}. The $+i\epsilon$ indicates the usual Feynman prescription for momentum-space propagators in Minkowski space in $(+--\hspace{0.09cm}-)$ signature, which we use throughout this work\footnote{In a direct position space interpretation of \eqref{2t} one should replace all four $+i\epsilon$'s by $-i\epsilon$'s.}. The integral (\ref{2t}) appears ubiquitously at next-to-leading order in the calculation of both scattering amplitudes and correlation functions in four-dimensional gauge theories. Concrete examples in planar $\mathcal{N}=4$ super-Yang-Mills (SYM) include the one-loop four gluon amplitude \cite{Bern_2005} and the one-loop correlation function of four protected operators \cite{Eden_1999}. In fact these two results are equal in a certain light cone limit \cite{Eden_2011}. More recently (\ref{2t}) has appeared directly as the first of a series of fishnet Feynman integrals in a strongly twisted version of $\mathcal{N}=4$ SYM \cite{Basso_2017}. It has been shown to possess a remarkable symmetry known as dual conformal invariance \cite{Broadhurst:1993ib,Drummond_2007}, which is best seen in Euclidean space. The Euclidean integral corresponding to (\ref{2t}) is 
\begin{align}\label{3t}
    I^E(x_1,x_2,x_3,x_4)=\int \frac{d^4 x_5}{\pi^2}\frac{x_{13}^2x_{24}^2}{x_{15}^2x_{25}^2x_{35}^2x_{45}^2},
\end{align}
where $x_1,\sp x_2,\sp x_3,\sp x_4$ are non-coincident points in Euclidean space $\mathbb{R}^4$. The dual conformal invariance of (\ref{3t}) refers to its invariance under conformal transformations in the $x$ variables
\begin{align}
    I^E(x_1,x_2,x_3,x_4)=I^E(x'_1,x'_2,x'_3,x'_4),\spac x_i'=Ax_i,
\end{align}
where $A:\mathbb{R}^4\rightarrow \mathbb{R}^4$ is a conformal transformation on $\mathbb{R}^4$, i.e.\ an element of $\text{Conf}(\mathbb{R}^4)\simeq SO(1,5)/\mathbb{Z}_2$. As such the Euclidean integral (\ref{3t}) depends only on a conformal invariant $z$ and its complex conjugate $\bar{z}=z^*$. It was calculated, for example in \cite{Usyukina:1992jd}, to be
\begin{align}\label{4t}
    I^E(x_1,x_2,x_3,x_4)=I^E(z)=\frac{2\text{Li}_2(z)-2\text{Li}_2(\bar{z})+\log(z\bar{z})\log(\frac{1-z}{1-\bar{z}})}{z-\bar{z}}.
\end{align}
Because $\bar{z}=z^*$ in Euclidean space (\ref{4t}) is essentially proportional to the Bloch-Wigner function, which has many interesting analytic properties \cite{Zagier:2007knq}. It is because of this conformal invariance that (\ref{3t}) is referred to as the \textit{dual conformal} box integral.
\newline
\newline
The situation in Minkowski space, where $z$ and $\bar{z}$ can either be independent real numbers or complex conjugate pairs, is more subtle. The dual conformal box integral in Minkowski space (\ref{2t}) has been calculated many times, first by 't Hooft and Veltman in \cite{tHooft:1978jhc}. More compact representations were obtained in \cite{Denner:1991qq} and \cite{Duplan_i__2002}. Common features of these calculations are the Feynman parametric and Mellin Barnes representations for the integral (\ref{2t}). The result in \cite{Duplan_i__2002} is the only one which gives any hint of the underlying dual conformal symmetry of the integral. It is given essentially in terms of the conformal invariants $z$ and $\bar{z}$, plus some corrections which depend explicitly on the signs of the kinematics\footnote{$x_{ij}^2$ coincide with the usual four particle kinematics $p_i^2,s,t$ in momentum space.} $x_{ij}^2$. These corrections encode a subtle breaking of dual conformal invariance in Minkowski space, which is caused by the Feynman $i\epsilon$, here serving as an IR regulator, required to properly define the integral. 
\newline
\newline
The $i\epsilon$ regulator encodes causality in the propagators, and the breaking of conformal invariance in Minkowski space is due to the fact that the conformal group on Minkowski space $\text{Conf}(\mathbb{R}^{1,3})\simeq SO^+(2,4)/\mathbb{Z}_2\simeq SU(2,2)/\mathbb{Z}_4$ does not preserve causality. In particular spacelike separations can be mapped to timelike separations, and vice versa. To define the action of the conformal group on Minkowski space properly, we must move to the compactification $\mathbb{R}_c^{1,3}$ \cite{book}. This compactification can be expressed in numerous ways, for example in terms of the group $U(2)$ \cite{Uhlmann1963THECO} or as the projective null cone in $\mathbb{R}^{2,4}$ \cite{Dirac:1936fq}. The conformal group then acts naturally on both of these spaces, although causality in the original Minkowski space is still not preserved. In order to preserve a notion of causality, it is necessary to move to the universal cover of compactified Minkowski space $\Tilde{\mathbb{R}}^{1,3}_c$ and consider the universal covering group $\reallywidetilde{\text{Conf}}(\mathbb{R}^{1,3})$ \cite{go1975}. In this work however we will focus on the conformal geometry of compactified Minkowski space, expressed in terms of $U(2)$. The value of the box integral (\ref{2t}) can change under finite special conformal transformations (SCTs) which change the signs of the kinematics $x_{ij}^2$. Geometrically this corresponds to a point `crossing infinity', i.e.\ moving from a point on Penrose's $\mathscr{I}^\pm$ to the antipodal point on $\mathscr{I}^{\mp}$. In the compactification these points are identified and crossing infinity still represents a continuous operation. For infinitesimal SCTs there are no issues and the integral (\ref{2t}) is \textit{locally} conformally invariant.
\newline
\newline
Another way to phrase the calculation of the Minkowskian box integral is ``how can one `analytically continue' the function (\ref{4t}) to Minkowski space?'' The Euclidean integral (\ref{3t}) can be obtained from (\ref{2t}) via a Wick rotation in certain kinematic regions, for example when all of the kinematics are spacelike ($x_{ij}^2<0$). In other regions (\ref{3t}) can only be Wick rotated to (\ref{2t}) up to logarithms of the conformal invariants $z$ and $\bar{z}$, so that the final answer is on another `branch' of the Bloch-Wigner function (\ref{4t}) \cite{hodges2010box}. There has been much work regarding the analytic continuation of Euclidean correlators to Minkowski space, and the subtleties of Wick rotation ~\cite{maldacena2015looking,Hartman_2016,Bautista_2020,qiao2020classification,cornalba2007eikonal}.
\newline
\newline
In this paper we classify all possible kinematic regions for four points in Minkowski space by their conformal structure, and describe how this relates to the branch jumping of the box integral (\ref{2t}). Specifically we study conformally equivalent configurations of four points in compactified Minkowski space. We find that, unlike in Euclidean space, the conformal invariants $z$ and $\bar{z}$ are not enough to determine when two configurations of four points are conformally equivalent. There are two cases, depending on whether $z$ and $\bar{z}$ are complex conjugates or real. If $z$ and $\bar{z}$ are complex conjugates, then all configurations with the same $z$ and $\bar{z}$ are conformally equivalent, as in the Euclidean case. If $z$ and $\bar{z}$ are real, then up to the discrete transformations of the Lorentz group $\mathcal{P},\sp\mathcal{T}$ there are two conformal equivalence classes of configurations. These two equivalence classes are exchanged by a reversal of the signs of each of the kinematics $x_{ij}^2$. In each conformal equivalence class we find that there are up to four values of the box integral (\ref{2t}), by explicitly rewriting the result of \cite{Duplan_i__2002}.
\newline
\newline
One new strategy for the calculation of (\ref{2t}) we explored is to exploit the richer structure of `infinity' in Minkowski space. In Euclidean space infinity is a single point $\infty$, whereas in compactified Minkowski space $\mathbb{R}^{1,3}_c$ it is a three-dimensional surface $\mathscr{I}\sp\cup\sp\iota$ (see section \ref{confcompact}). All configurations of four points in Minkowski space can be conformally mapped to a configuration with two or more points at infinity, such that (importantly) the value of the box integral does not change. Most configurations with real $z,\sp\bar{z}$ can be conformally mapped to a special `double infinity' configuration, where the box integral is calculable directly in spacetime using only the residue theorem. The rest of the configurations can be mapped to a configuration with three or four points at infinity, however these are not discussed in this paper as the corresponding integrals are not as easily calculated. These higher infinity configurations, similarly to $(0,1,z,\infty)$ in Euclidean space, furnish a `minimal' set of configurations which fully fix the box integral (and other four-point observables) by pseudo-conformal invariance. They also provide a nice way to generate configurations of four points with a given $z,\bar{z}$, in a given kinematic region. 
\newline
\newline
It is worth noting that the box integral has been shown to be Yangian invariant \cite{Chicherin_2017,Chicherin_2018}, and in Euclidean space it has been successfully constrained to the Bloch-Wigner function using the novel `Yangian bootstrap' approach \cite{Loebbert_2020,loebbert2020yangian}. The Minkowskian box integral has also been considered in a momentum-twistor geometry by Hodges \cite{hodges2010box}, who also discussed the subtle breaking of dual conformal invariance and analytic continuation to all kinematic regions of Minkowski space.
\newline
\newline
The structure of this paper is as follows. In section \ref{confcompact} we review the action of the conformal group on the $U(2)$ conformal compactification of Minkowski space. In section \ref{confplane} we study the conformal geometry of four points in compactified Minkowski space, and describe the sets of conformally equivalent configurations. In section \ref{symmetry} we discuss the symmetries of the box integral directly in Minkowski space, and the various representations of the integral which make these symmetries manifest. We give an expression for the box integral (\ref{2t}) in all kinematic regions by specialising the result of \cite{Duplan_i__2002}. In section \ref{doubleinf} we discuss the calculation of the box integral in the special `double infinity' configurations. We found the package \cite{vanOldenborgh:1990yc} useful for numerically testing calculations. We always take logarithms with a branch cut on the negative real axis, so that the corresponding dilogarithms have a cut on $(1,\infty)$.

\section{Conformally Compactified Minkowski Space}\label{confcompact}
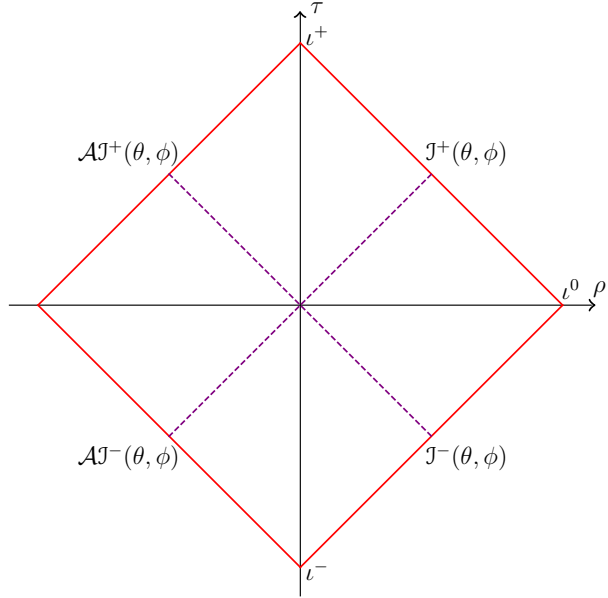
\begin{figure}[h!]
\begin{center}
\begin{tikzpicture}[thick,scale=0.63, every node/.style={scale=0.785}]

\node[inner sep=0pt] (russell) at (0,0)
    {\includegraphics[width=.6\textwidth]{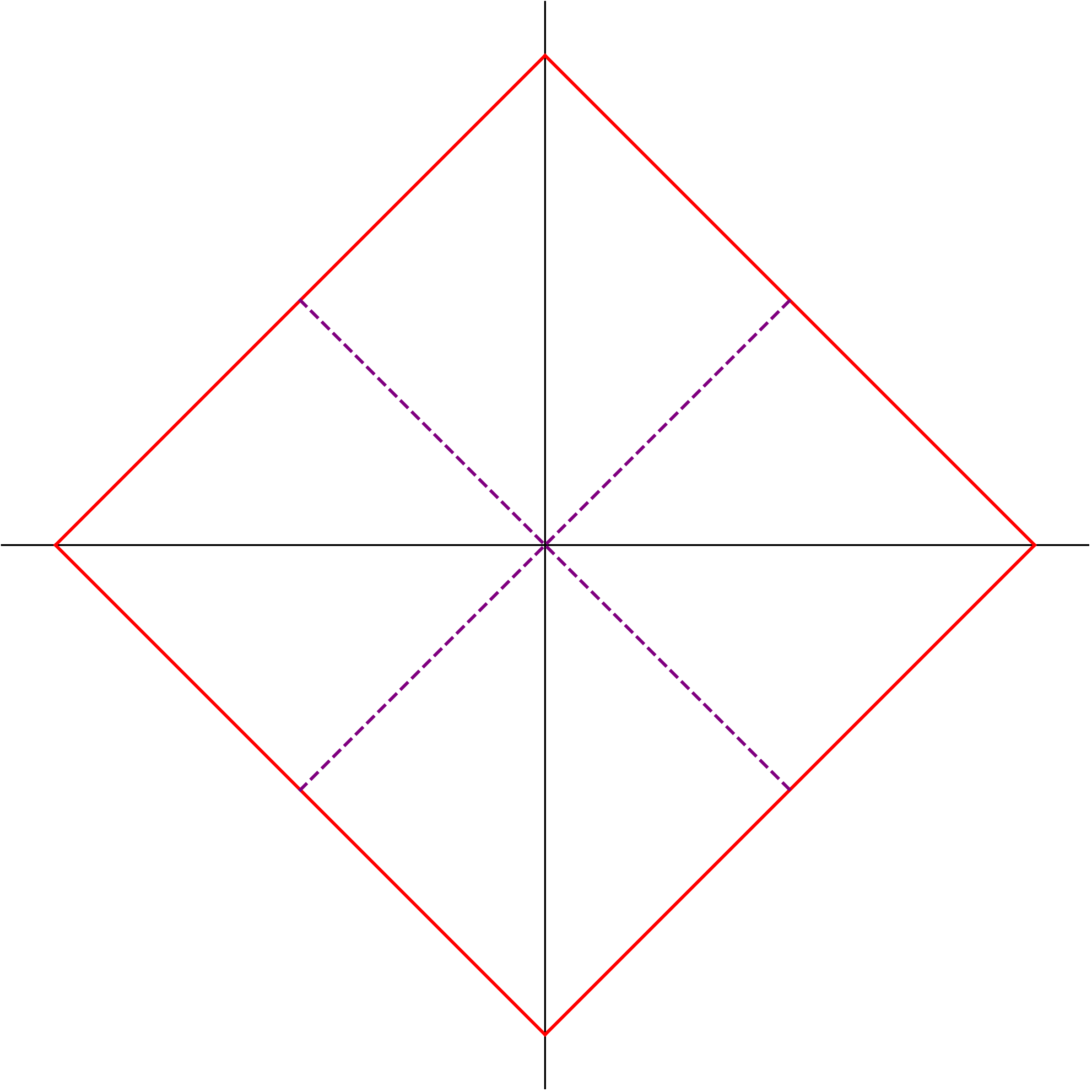}};
    \draw[->,thick] (-.015,6.2)--(-0.015,6.22);
 \draw[->,thick] (6.2,0.01)--(6.22,0.01);
 \node at (6.3,0.35){$\rho$};
  \node at (0.35,6.3){$\tau$};
   \node at (0.35,5.7){$\iota^+$};
     \node at (0.35,-5.6){$\iota^-$};
       \node at (5.7,0.35){$\iota^0$};
     \node at (3.5,3.2){$\mathscr{I}^+(\theta,\phi)$};
       \node at (3.5,-3.2){$\mathscr{I}^-(\theta,\phi)$};
        \node at (-3.63,3.2){$\mathcal{A}\mathscr{I}^+(\theta,\phi)$};
       \node at (-3.63,-3.2){$\mathcal{A}\mathscr{I}^-(\theta,\phi)$};
\end{tikzpicture}
\end{center}
    \caption{Extended Penrose diagram of Minkowski space. Every two-sphere $S^2(\tau,\rho>0)$ is represented by two points, one on the left and one on the right, which are exchanged by the antipodal map $\mathcal{A}$. The dotted lines represent the light cone of the origin \cite{strominger2017lectures}.}
    \label{figg11}
\end{figure}

\begin{figure}[h!]
\begin{center}
\begin{tikzpicture}[thick,scale=1, every node/.style={scale=0.81}]

  \draw[thick] (1.33,0) ellipse (1.3 and 1);

\draw[thick] (0,0) arc (180:-180:2);

\draw[blue,thick]    (2.62,0) to[out=-50,in=-130] (4,0);
\draw[blue,dotted]    (2.62,0) to[out=50,in=130] (4,0);

\draw[red,thick]    (2.1,0.8) to[out=0,in=-20] (3.05,1.7);
\draw[red,dotted]    (3.05,1.7) to[out=180,in=160] (2.1,0.8);
\draw[red,dotted]    (2.1,-0.8) to[out=0,in=20] (3.05,-1.7);
\draw[red,thick]    (3.05,-1.7) to[out=180,in=-160] (2.1,-0.8);

\draw[red,thick]    (1.3,1) to[out=10,in=-10] (1.2,1.85);
\draw[red,dotted]    (1.3,1) to[out=170,in=190] (1.2,1.85);

\draw[red,dotted]    (1.3,-1) to[out=10,in=-10] (1.2,-1.85);
\draw[red,thick]    (1.3,-1) to[out=170,in=140] (1.2,-1.82);

\filldraw[black] (0.02,0) circle (1pt);
\footnotesize
\node at (-0.6,0){$\iota:\chi=0$};
\node at (4.7,0){$\chi=\pi/2$};
\node at (3.47,1.87){$\chi=\frac{\pi}{3}$};
\node at (3.49,-1.85){$\chi=-\frac{\pi}{3}$};
\node at (1.1,2.1){$\chi=\frac{\pi}{6}$};
\node at (1.2,-2.17){$\chi=-\frac{\pi}{6}$};
\node at (3.35,0.0){\textcolor{blue}{$S^*$}};
\normalsize
    \node at (2.25,2.2){$\mathscr{I}$};
\end{tikzpicture}
\end{center}
    \caption{Conformal infinity in terms of $\iota$ and $\mathscr{I}$. Surfaces of fixed $\chi$ are $2-$spheres. The parameter $\chi\in(-\pi/2,\pi/2]$ is chosen as $\chi=\pi/2-\chi_+$ mod $\pi$, so $\chi=0$ corresponds to $\iota$.}
    \label{figg12}
\end{figure}
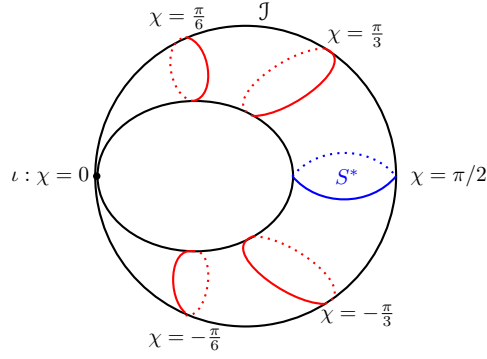

There is a slight subtlety in defining the action of the conformal group on Euclidean space $\mathbb{R}^4$. Since the inversion operation $\mathcal{I}:x\rightarrow x/x^2$ is not well-defined on the origin $x=0$, we should add a point `$\infty$' to $\mathbb{R}^4$ which corresponds to the image of the origin under $\mathcal{I}$. The conformal group is well-defined and acts transitively\footnote{A group $G$ acts transitively on a set $X$ if for all $x,y\in X$ there exists $g\in G$ such that $gx=y$.} on the \textit{compactification} $\mathbb{R}_c^4\equiv \mathbb{R}^4\cup\{\infty\}$.
\newline
\newline
The situation is more subtle in Minkowski space $\mathbb{R}^{1,3}$. Defining the light cone of the origin
\begin{align}
    L_0\equiv \{x\in\mathbb{R}^{1,3}\sp\sp|\sp\sp x^2=0\}
\end{align}
we note in this case that $\mathcal{I}$ not well-defined on the entire space $L_0$. We thus are interested in a compactification $\mathbb{R}^{1,3}\rightarrow \mathbb{R}_c^{1,3}$ such that the conformal group is well-defined and acts transitively and continuously on $\mathbb{R}_c^{1,3}$. The points `at infinity' will be identifiable\footnote{Up to an extra `null sphere at infinity' $S^*$, see (\ref{u4}) and figure \ref{figg12}.} with the image $\mathcal{I}(L_0)$, and so will constitute a $3-$dimensional surface.
\newline
\newline
The most common way of compactifying Minkowski space $(\mathbb{R}^{1,3},\eta)$ is the \textit{Penrose compactification} \cite{Penrose:1964ge} and replaces $\mathbb{R}^{1,3}\rightarrow \mathbb{R}^{1,3}_p\equiv \mathbb{R}^{1,3}\cup \partial \mathbb{R}^{1,3}_p$, where $\partial \mathbb{R}^{1,3}_p\equiv\iota^+\cup\iota^-\cup\iota^0\cup\mathscr{I}^+\cup\mathscr{I}^- $. $\iota^0$ is called \textit{spatial} infinity, and in spherical coordinates $(t,r,\theta,\phi)$ corresponds to $r\rightarrow \infty$. $\iota^\pm$ are called \textit{future} and \textit{past} infinity, and correspond to $t\rightarrow \pm\infty$. Both $\iota^0$ and $\iota^\pm$ correspond to individual points on $\partial\mathbb{R}_p^{1,3}$. $\mathscr{I}^\pm$ are called \textit{future null} and \textit{past null} infinity and correspond to $t\pm r\rightarrow \infty$ with $t\mp r$ constant. $\mathscr{I}^{\pm}$ are $3-$dimensional submanifolds of $\partial\mathbb{R}_p^{1,3}$, parametrised by the coordinates $(t\mp r,\theta,\phi)$. Useful coordinates for describing $\mathbb{R}^{1,3}_p$ are
\begin{align}
\chi_+\equiv \arctan(t+r),\spac \chi_-\equiv\arctan(t-r), 
\end{align}
and $\tau\equiv \chi_-+\chi_+,\sp\sp\sp \rho\equiv\chi_+-\chi_-$. The range of the $\tau,\sp \rho$ coordinates in $\mathbb{R}^{1,3}_p$ is $\tau\in[-\pi,\pi],\sp\rho\in[0,\pi]$. We visualise $\mathbb{R}^{1,3}_p$ in $(\tau,\rho)$ coordinates using an extended Penrose diagram (figure \ref{figg11}).
\newline
We want to consider conformal transformations on the Penrose compactification $\mathbb{R}_p^{1,3}$. There are a few issues at $\partial \mathbb{R}_p^{1,3}$. Using infinitesimal SCTs (\ref{u3}) it is possible to move between $\iota^\pm$ and $\iota^0$, and also between $\mathscr{I}^{\pm}(\chi_\mp,\theta,\phi)$ and $\mathscr{I}^{\mp}(\chi_\pm,\mathcal{A}(\theta,\phi))$, where $\mathcal{A}$ is the antipodal map on $S^2$. Therefore in order for the conformal group to act continuously on compactified Minkowski space, we must make the identifications\footnote{The identification $\mathscr{I}^+=\mathcal{A}\mathscr{I}^-$ resembles boundary conditions on fields in asymptotic symmetries \cite{strominger2017lectures}.}
\begin{align}\label{3r}
\iota^+=\iota^-=\iota^0\equiv\iota,\spac \mathscr{I}^+=\mathcal{A}\mathscr{I}^-\equiv \mathscr{I}. 
\end{align}
We define $\partial \mathbb{R}^{1,3}_c\equiv \iota\sp\cup\sp\mathscr{I}$ as $\partial \mathbb{R}_p^{1,3}$ subject to the identifications (\ref{3r}). $\partial \mathbb{R}_c^{1,3}$ can be visualised as a `pinched torus' or croissant (figure \ref{figg12}), and can be parametrised in terms of $\chi\in(-\pi/2,\pi/2]$ and a 2$-$sphere angle $(\theta,\phi)$ for $\chi\neq 0$ \cite{Jadczyk_2011}. We define the conformal compactification $\mathbb{R}^{1,3}_c\equiv \mathbb{R}^{1,3}\cup \partial \mathbb{R}_c^{1,3}$. $\mathbb{R}^{1,3}_c$ can be identified with the group $U(2)$, on which the conformal group acts naturally by $SU(2,2)/\mathbb{Z}_4$.
To see this, we first note the bijection $H:\mathbb{R}^{1,3}\rightarrow H_{2\times2}$ between Minkowski space and the space of $2\times 2$ Hermitian matrices
\begin{align}
    H:x=(x^0,x^1,x^2,x^3)\longrightarrow H(x)\equiv\begin{pmatrix}x^0+x^3&x^1-ix^2\\ x^1+ix^2&x^0-x^3\end{pmatrix}.
\end{align}
In what follows we will label the Hermitian matrices corresponding to points $x,y,z$ in Minkowski space by $X,Y,Z$.
The inverse map is given by
\begin{align}\label{30a}
    H^{-1}:X\rightarrow x^\mu=\frac{1}{2}\text{tr}(X\sigma^\mu),
\end{align}
where $\si^{\mu}=(\mathbb{I}_2,\si^1,\si^2,\si^3)$. In spherical coordinates $x=(t,r\sin\theta\cos\phi,r\sin\theta\sin\phi,r\cos\theta)$, the Hermitian matrix corresponding to $x$ is
\begin{align}
    X=\begin{pmatrix}t+r\cos\theta& re^{-i\phi}\sin\theta\\ re^{i\phi}\sin\theta&t-r\cos\theta\end{pmatrix}=\Omega(\theta,\phi)^{\dagger}M(t,r)\Omega(\theta,\phi),
\end{align}
where
\begin{align}\label{8r}
\Omega(\theta,\phi)\equiv\begin{pmatrix}\cos(\theta/2)& e^{-i\phi}\sin(\theta/2)\\ -e^{i\phi}\sin(\theta/2)&\cos(\theta/2)\end{pmatrix}\in SU(2),\spac M(t,r)\equiv \begin{pmatrix}t+r&0\\ 0&t-r\end{pmatrix}.
\end{align}
$M(t,r)$ represents the event in Minkowski space at a time $t$, and spatially on the north pole ($\theta=0$) of a sphere of radius $r$. The adjoint action of $\Omega(\theta,\phi)$ rotates this point from the north pole to a general angle $(\theta,\phi)$ on the sphere. In terms of Hermitian matrices we can calculate spacetime intervals via determinants
\begin{align}
    (x-y)^2=\det(X-Y).
\end{align}
To compactify, we use the injective Cayley map\footnote{This is analogous to the one-dimensional Cayley map $\mathbb{R}\rightarrow U(1)$, $x\rightarrow \frac{i-x}{i+x}$} $J:H_{2\times 2}\rightarrow U(2)$ defined by
\begin{align}\label{1j}
    J:X\longrightarrow  J(X)\equiv \frac{i\mathbb{I}_2-X}{i\mathbb{I}_2+X},
\end{align}
which is well-defined because $\det(i\mathbb{I}_2+X)\neq 0$ and $[i\mathbb{I}_2-X,(i\mathbb{I}_2+X)^{-1}]=0$ for all $X\in H_{2\times 2}$. The map (\ref{1j}) is not surjective. The inverse is given
\begin{align}
    J^{-1}:U\rightarrow i\frac{\mathbb{I}_2-U}{\mathbb{I}_2+U},
\end{align}
which is not well-defined for those $U\in U(2)$ which satisfy $\det(\mathbb{I}_2+U)=0$. Im$(J)$ can be identified with Minkowski space $\mathbb{R}^{1,3}$ and $U_\infty\equiv U(2)\setminus \text{Im}(J)=\{U\in U(2)\sp|\sp \det(\mathbb{I}_2+U)=0\}$ can be identified with the conformal boundary $\partial \mathbb{R}_c^{1,3}$ defined above. If $x_1$ and $x_2$ are points in Minkowski space represented by unitary matrices $U_1$ and $U_2$, their spacetime interval can be calculated
\begin{align}\label{25j}
    (x_1-x_2)^2=-4\frac{\det(U_1-U_2)}{\det(\mathbb{I}_2+U_1)\det(\mathbb{I}_2+U_2)}.
\end{align} 
We can define the spacetime interval between arbitrary unitary matrices $U_1,U_2$ using the formula (\ref{25j}), noting that the expression is of course infinite if either $U_1$ or $U_2$ are on $U_{\infty}$ so that the interval is not well-defined. To define a finite interval between all unitary matrices an unphysical metric $\langle U_1,U_2\rangle\equiv \det(U_1-U_2)$ must be used. It is however possible to define conformally invariant combinations of any four unitary matrices $U_1,U_2,U_3,U_4$ using the physical metric, as seen in section \ref{confplane}.
\newline
\newline
Conformal transformations can be implemented on $\mathbb{R}^{1,3}_c\simeq U(2)$ by $SU(2,2)$ fractional linear transformations \cite{Jadczyk_2011,1976JMP....17...24P}
\begin{align}
      U\longrightarrow \mathfrak{C}_GU\equiv (AU+B)(CU+D)^{-1},\spac G=\begin{pmatrix}A&B\\C&D\end{pmatrix}\in SU(2,2),
\end{align}
\begin{align}
    SU(2,2)=\{G\in SL(4,\mathbb{C})\sp\sp|\sp\sp G^{\dagger}KG=K\},\spac K=\begin{pmatrix}\mathbb{I}_2&0\\0&-\mathbb{I}_2\end{pmatrix}.
\end{align}
$\mathfrak{C}$ is the fourfold covering homomorphism $\mathfrak{C}: SU(2,2)\rightarrow\text{Conf}(\mathbb{R}^{1,3})$. $\mathfrak{C}$ acts trivially on the centre $Z(SU(2,2))=\{\pm\mathbb{I}_4,\pm i\mathbb{I}_4\}\simeq \mathbb{Z}_4$ and $\text{Conf}(\mathbb{R}^{1,3})\simeq SU(2,2)/ \mathbb{Z}_4$. The $SU(2,2)$ matrices which correspond to the familiar conformal transformations on Minkowski space are given in appendix \ref{su22}. The set $U_\infty$ can be parametrised explicitly
\begin{align}
    U_{\infty}=\left\{\Omega(\theta,\phi)^{\dagger}\begin{pmatrix}-1 & 0\\0&-e^{2i\chi}\end{pmatrix}\Omega(\theta,\phi)\in U(2)\spa\Big|\spa \chi\in(-\pi/2,\pi/2]\right\},
\end{align}
where $\Omega(\theta,\phi)$ is as defined in (\ref{8r}). With the $U(2)$ formalism it can be checked that $\partial \mathbb{R}_c^{1,3}=U_{\infty}$ and $U_{\infty}=\mathcal{I}(L_0)\cup S^*$. $S^*$ is the \textit{null sphere at infinity} and can be parametrised
\begin{align}\label{u4}
   S^*=\left\{\Omega(\theta,\phi)^{\dagger}\begin{pmatrix}-1 & 0\\0&1\end{pmatrix}\Omega(\theta,\phi)\in U(2)\right\}.
\end{align}
$S^*$ is stable under inversions $\mathcal{I}S^*=S^*$. It corresponds to the largest sphere ($\chi=\pi/2$) on the pinched torus (figure \ref{figg12}) and can be reached from the bulk of Minkowski space by taking limits to infinity along the light cone $L_0$. $\iota$ is the image of the origin under $\mathcal{I}$ and corresponds to the $U(2)$ matrix $U=-\mathbb{I}_2$.

\section{Conformal Cross-ratios and Conformal Planes}\label{confplane}
The group of all conformal transformations on compactified Minkowski space $\mathbb{R}^{1,3}_c$ is isomorphic to $O(2,4)/\mathbb{Z}_2$. It is generated by translations, rotations $R\in O(1,3)$, dilatations, and inversions $\mathcal{I}$. The \textit{conformal group} $\text{Conf}(\mathbb{R}^{1,3})$ is defined to be the connected component of the identity of this group, and is isomorphic to $SO^+(2,4)/\mathbb{Z}_2\simeq SU(2,2)/\mathbb{Z}_4$ \cite{book}. We will use the connected component $\text{Conf}(\mathbb{R}^{1,3})$, and explicitly refer to the extra discrete transformations such as $\mathcal{P}$ and $\mathcal{T}$ when needed. For $N\geq 4$ points $U_1, U_2,\dots, U_N\in\mathbb{R}^{1,3}_c$, expressed as unitary matrices, one can form $\frac{N}{2}(N-3)$ independent conformally invariant combinations known as \textit{conformal cross-ratios}. Here we are interested in $N=4$ points, where given $U_1,U_2,U_3,U_4\in\mathbb{R}^{1,3}_c$ the four point cross-ratios $u$ and $v$ can be defined
\begin{align}\label{10j}
    u=\frac{\det(U_1-U_2)\det(U_3-U_4)}{\det(U_1-U_3)\det(U_2-U_4)},\spac  v=\frac{\det(U_1-U_4)\det(U_2-U_3)}{\det(U_1-U_3)\det(U_2-U_4)}.
\end{align}
If none of the points are on $U_\infty$ then the points can be expressed as usual Minkowski vectors $x_1,x_2,x_3,x_4$ and the cross-ratios can be calculated
\begin{align}\label{9j}
    u=\frac{x_{12}^2x_{34}^2}{x_{13}^2x_{24}^2},\spac  v=\frac{x_{14}^2x_{23}^2}{x_{13}^2x_{24}^2}.
\end{align}
If one or more points are on $U_\infty$, then the cross-ratios can be calculated with (\ref{10j}) or by taking appropriate limits of (\ref{9j}). We define $V$ to be the set of $4-$tuples $\boldsymbol{U}=(U_1,U_2,U_3,U_4)$ of points in $\mathbb{R}^{1,3}_c$, such that the cross-ratios are non-singular\footnote{If none of the points are on $U_\infty$ this corresponds to none of the points being lightlike separated.} 
\begin{align}\label{11j}
   V=\left\{\boldsymbol{U}=(U_1,U_2,U_3,U_4)\spa\Big|\spa U_i\in\mathbb{R}^{1,3}_c,\spa u(\boldsymbol{U}),\sp v(\boldsymbol{U})\neq 0,\sp\infty\right\} .
\end{align} 
Non-singularity of the cross-ratios implies the determinants $\det(U_i-U_j)$ are nonvanishing for $i<j$. Points $\boldsymbol{U}\in V$ will be referred to as \textit{configurations} of $\mathbb{R}^{1,3}_c$. If none of the points in $\boldsymbol{U}\in V$ are on $U_{\infty}$, then we call $\boldsymbol{U}\equiv\boldsymbol{w}$ a \textit{finite} configuration. A finite configuration $\boldsymbol{w}\in V$ can be written in terms of usual Minkowski vectors $\boldsymbol{w}=(x_1,x_2,x_3,x_4)$ with $x_i\in\mathbb{R}^{1,3}$. We denote the set of finite configurations by $V_f\subset V$. The cross-ratios $u$ and $v$ are commonly expressed in terms of other conformally invariant quantities $z$ and $\bar{z}$, defined by
\begin{align}
    u=z\bar{z},\spac v=(1-z)(1-\bar{z}).
\end{align}
Note that since we defined $u,v\neq 0$ in (\ref{11j}), for configurations $\boldsymbol{U}\in V$ we have that $z,\bar{z}\neq 0,1$. $z$ and $\bar{z}$ can be extracted from $u$ and $v$ by the formula 
\begin{align}\label{2h}
    z,\bar{z}=\frac{1+u-v}{2}\pm\sqrt{\left(\frac{1-u-v}{2}\right)^2-uv}.
\end{align}
There are two things to note from (\ref{2h}). The first is that $z$ and $\bar{z}$ can in general be either real numbers or complex conjugate pairs. Secondly since $z$ and $\bar{z}$ appear as the roots of a quadratic equation, without loss of generality we can (and will) always take $\bar{z}\geq z$ when $z,\bar{z}\in\mathbb{R}$ and $\text{Im}(z)>0$ when $z,\bar{z}\in \mathbb{C}\setminus\mathbb{R}$. This is not necessary but we find it useful in practice. In what follows we will refer to $z$ and $\bar{z}$ for a given configuration as the \textit{conformal invariants} for that configuration. Define $V_{z,\bar{z}}$ as the subset of $V$ such that $\boldsymbol{U}=(U_1,U_2,U_3,U_4)\in V$ has conformal invariants $z,\bar{z}$, i.e.\
\begin{align}\label{12j}
    V_{z,\bar{z}}=\left\{\boldsymbol{U}\in V\spa\spa \Big|\spa\spa u(\boldsymbol{U})=z\bar{z},\spa v(\boldsymbol{U})=(1-z)(1-\bar{z})\right\}.
\end{align}
Based on the possible values for $z$ and $\bar{z}$, we have the decomposition
\begin{align}\label{3h}
    V=\bigcup_{\substack{z,\bar{z}\in \mathbb{R}\setminus \{0,1\}\\z\leq\bar{z}}}V_{z,\bar{z}}\cup\bigcup_{\substack{z\in \mathbb{H}\\\bar{z}=z^*}} V_{z,\bar{z}}\equiv V_{\mathbb{R}}\cup V_{\mathbb{C}},
\end{align}
where $\mathbb{H}\subset \mathbb{C}$ is the upper half-plane. Note that the `boundary' between $V_{\mathbb{R}}$ and $V_{\mathbb{C}}$ occurs when $z=\bar{z}$, and is included in $V_{\mathbb{R}}$. If $A:\mathbb{R}^{1,3}_c\rightarrow \mathbb{R}^{1,3}_c$ is a conformal transformation and $\boldsymbol{U}=(U_1,U_2,U_3,U_4)\in V$, we define $A$ to act on $\boldsymbol{U}$ by
\begin{align}
    A:\boldsymbol{U}\rightarrow A\boldsymbol{U}\equiv(AU_1,\sp AU_2,\sp AU_3,\sp AU_4).
\end{align}
There are a number of questions one can ask about the decomposition (\ref{3h}). Ranging over configurations $\boldsymbol{U}\in V$ what are the possible values of $z$ and $\bar{z}$, i.e.\ which of the $V_{z,\bar{z}}$ are non-empty? Furthermore does the conformal group $\text{Conf}(\mathbb{R}^{1,3})$ act transitively on each $V_{z,\bar{z}}$? These questions have been studied in the case of null hexagon configurations in compactified Minkowski space \cite{dorn2012conformal}. We will briefly review these questions for Euclidean space before moving to Minkowski space.

\subsection{Euclidean Space}\label{confplaneeuc}
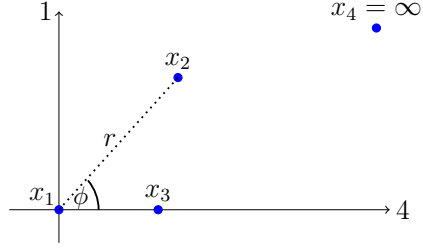
\begin{figure}[h!]
\begin{center}
\begin{adjustbox}{max totalsize={.34\textwidth}{.27\textheight},center}
\begin{tikzpicture}{scale=0.2} 

\draw[->] (-0.75,0)--(5,0); 
\draw[->] (0,-0.5)--(0,3); 
\node at (-0.2,3){$1$};
\node at (5.2,0){$4$};
\node at (-0.25,0.2){$x_1$};
\filldraw[blue] (0,0) circle (1.75pt);
\node at (1.5,0.25){$x_3$};
\filldraw[blue] (1.5,0) circle (1.75pt);
\filldraw[blue] (1.8,2) circle (1.75pt);
\draw[dotted,thick] (0,0)--(1.8,2);
\node at (1.8,2.25){$x_2$};
\node at (0.77,1.05){$r$};
\filldraw[blue] (4.8,2.75) circle (1.75pt);
\node at (4.8,3){$x_4=\infty$};
\small
\node at (0.33,0.18){$\phi$};
\draw[thick] (0.6,0) arc (0:40:0.69);
\normalsize
\end{tikzpicture}
\end{adjustbox}
\end{center}
\caption{Euclidean conformal plane configuration $\bar{w}(r,\phi)$.}
    \label{figg19}
\end{figure}

We consider compactified Euclidean space $\mathbb{R}^{4}_c=\mathbb{R}^4\cup\{\infty\}$. We define $V^E$ to be the set of $4-$tuples of non-coincident points in $\mathbb{R}^{4}_c$, and $V^E_{z,\bar{z}}$ to be the subset of these with conformal invariants $z$ and $\bar{z}$, in analogy with the Minkowskian definitions (\ref{11j}) and (\ref{12j}). In this case it is well-known that for any configuration $\boldsymbol{w}=(x_1,x_2,x_3,x_4)\in V^E$ we have that $\bar{z}=z^*$ and that $\text{Conf}(\mathbb{R}^{4})$ acts transitively on each $V^E_{z,\bar{z}}$.
\newline
\newline
Indeed we can map any $\boldsymbol{w}=(x_1,x_2,x_3,x_4)\in V^{E}$ into a canonical Euclidean configuration $\bar{\boldsymbol{w}}(r,\phi)\in V^{E}$, defined below, as follows. We assume the points are generic so that none of $x_1,x_2,x_3,x_4$ are $\infty$. We make the conformal transformation $A_1\equiv T_{-\mathcal{I}x_1}\sp\sp\mathcal{I}\sp\sp T_{-x_4}$. Under $A_1$ $\boldsymbol{w}$ transforms
\begin{align}
    \boldsymbol{w}\rightarrow \boldsymbol{w}_1\equiv A_1\boldsymbol{w}=(0,y_2,y_3,\infty),
\end{align}
where $y_2\equiv A_1x_2, y_3\equiv A_1x_3$. The stabilising conformal subgroup of the points $0$ and $\infty$ consists of dilatations and rotations $SO(4)\subset SO(1,5)$. Let $|y_3|^2=\rho^2$. Then we can pick $R\in SO(4)$ to rotate $y_3\rightarrow (0,0,0,\rho)$ and then rescale by $1/\rho$ to reach the unit vector $e_4=(0,0,0,1)$. So we let $A_2\equiv D_{1/\rho}R$ and 
\begin{align}
    \boldsymbol{w}_1\rightarrow \boldsymbol{w}_2\equiv A_2\boldsymbol{w}_1=(0,\sp\sp z_2,\sp\sp e_4,\sp\sp\infty),
\end{align}
where $z_2\equiv A_2y_2$. The stabiliser of $0$, $e_4$, and $\infty$ is $SO(3)\subset SO(4)$ acting on the first 3 coordinates of $\mathbb{R}^4$. We pick $A_3\equiv R'\in SO(3)$ which maps $z_2$ to the 14 plane $z_2\rightarrow (r\sin\phi,0,0,r\cos\phi)$, with $r>0$ and $\phi\in [0,2\pi)$. Defining $A\equiv A_3A_2A_1$ we have shown that given any $\boldsymbol{w}\in V^E$
\begin{align}
  \bar{\boldsymbol{w}}(r,\phi)\equiv   A\boldsymbol{w}= (0,\sp\sp (r\sin\phi,0,0,r\cos\phi),\sp\sp e_4,\sp\sp\infty)
\end{align}
for some $r>0, \phi\in[0,2\pi)$. We call $\bar{\boldsymbol{w}}(r,\phi)$ the \textit{Euclidean conformal plane configuration} (figure \ref{figg19}).
\newline\newline
The conformal invariants of $\bar{\boldsymbol{w}}(r,\phi)$, and hence of $\boldsymbol{w}$, are calculated to be
\begin{align}
    z=re^{i\phi},\spac \bar{z}=re^{-i\phi}.
\end{align}
If $\phi\in (\pi,2\pi)$ we can relabel $\phi\rightarrow 2\pi-\phi$ so that $\text{Im}(z)\geq0$. Therefore we can always take $\phi\in[0,\pi]$. If $\phi=0$ or $\pi$ then $z=\bar{z}\in \mathbb{R}\setminus \{0,1\}$. Therefore the decomposition (\ref{3h}) for Euclidean space reads
\begin{align}
    V^E=\bigcup_{\substack{z=\bar{z}\in \mathbb{R}\setminus \{0,1\}}}V^E_{z,\bar{z}}\cup\bigcup_{\substack{z\in \mathbb{H}\\\bar{z}=z^*}} V^E_{z,\bar{z}}\equiv V_{\mathbb{R}}^E\cup V_{\mathbb{C}}^E.
\end{align}
Does $\text{Conf}(\mathbb{R}^4)$ act transitively on each $V^E_{re^{i\phi},re^{-i\phi}}$? The answer is easily yes. Given $\boldsymbol{w},\boldsymbol{y}\in V^E_{re^{i\phi},re^{-i\phi}}$, we can always find conformal transformations $A_w,A_y$ which map $\boldsymbol{w}$ and $\boldsymbol{y}$ to the conformal plane $A_w\boldsymbol{w}=A_y\boldsymbol{y}=\bar{\boldsymbol{w}}(r,\phi)$. Then if we let $A\equiv A_w^{-1}A_y$ we see that $\boldsymbol{w}=A\boldsymbol{y}$.

\subsection{Minkowski Space}\label{confplanemink}
The situation is much more subtle in compactified Minkowski space $\mathbb{R}^{1,3}_c$, where the squared differences $x_{ij}^2$ between finite points can be positive or negative. The conformal structure is much richer, and we will see that $z$ and $\bar{z}$ can be complex conjugate pairs, or independent real numbers. Given a finite configuration $\boldsymbol{w}=(x_1,x_2,x_3,x_4)\in V_f$, we define the \textit{kinematics} $k(\boldsymbol{w})$ of $\boldsymbol{w}$ to be the $6-$tuple
\begin{align}
   k(\boldsymbol{w})\equiv (x_{12}^2,x_{34}^2,x_{23}^2,x_{14}^2,x_{13}^2,x_{24}^2).
\end{align}
As a slight abuse of terminology, we will also refer to the individual $x_{ij}^2$ as kinematics. We restrict to finite configurations $\boldsymbol{w}$ because the kinematics are not well-defined when at least one of the points is on $U_{\infty}$. We can also take a generic configuration to be finite. We define the \textit{sign} of the kinematics (and similarly for the sign of any ordered tuple)
\begin{align}
    \text{sgn}(k(\boldsymbol{w}))\equiv (\sp\text{sgn}(x_{12}^2),\spa\text{sgn}(x_{34}^2),\spa\text{sgn}(x_{23}^2),\spa\text{sgn}(x_{14}^2),\spa\text{sgn}(x_{13}^2),\spa\text{sgn}(x_{24}^2)\sp),
\end{align}
where sgn$(x)=+$ if $x>0$ and sgn$(x)=-$ if $x<0$. By non-singularity of the cross-ratios (\ref{11j}) none of the squared differences for finite configurations vanish and so sgn$(x_{ij}^2)\in\{-,+\}$ for each $i<j$. Therefore depending on $\boldsymbol{w}\in V_f$ there are $2^6=64$ possibilities for sgn$(k(\boldsymbol{w}))$. We will concatenate signs as a shorthand, so for example sgn$(1,-3,2,5,5,2)=(+,-,+,+,+,+)\equiv (+-+++\hspace{0.02cm}+)$. We denote by $P_{ij}$ the operator which flips the sign of $x_{ij}^2$, so for example $P_{13} (+++++\hspace{0.02cm}+)=(++++-\hspace{0.02cm}+)$. Moreover we define $P\equiv \prod_{i<j}P_{ij}$ as the operator which flips the sign of all kinematics, so for example $P(+-+-+\hspace{0.05cm}-)=(-+-+-\hspace{0.05cm}+)$. 
\newline
\newline
We again ask the question: Does $\text{Conf}(\mathbb{R}^{1,3})$ act transitively on each $V_{z,\bar{z}}$? In this case the answer is \textit{no} in general, and the existence of $A\in \text{Conf}(\mathbb{R}^{1,3})$ connecting finite configurations $\boldsymbol{w},\boldsymbol{y}\in V_{z,\bar{z}}$ depends on $k(\boldsymbol{w})$ and $k(\boldsymbol{y})$. Inversions and SCTs can change the kinematics of a configuration, in such a way that the cross-ratios are fixed. However they can only change the \textit{signs} of the kinematics in a restricted way. For the rest of this section we refine (\ref{3h}) such that $\text{Conf}(\mathbb{R}^{1,3})$ acts transitively on each set in the decomposition. This is possible provided the discrete transformations $\mathcal{P},\mathcal{T}$ are also used. 
\newline
\newline
Let $b\in\mathbb{R}^{1,3}$, $\boldsymbol{w}=(x_1,x_2,x_3,x_4)\in V_f$. Under a special conformal transformation $\boldsymbol{w}\rightarrow C_b\boldsymbol{w}$ the kinematics transform
\begin{align}\label{10f}
    x_{ij}^2\rightarrow \frac{x_{ij}^2}{\si_b(x_i)\si_b(x_j)},\spac \text{sgn}(x_{ij}^2) \rightarrow \frac{\text{sgn}(x_{ij}^2)}{\text{sgn}(\si_b(x_i)\si_b(x_j))},
\end{align}
where $\si_b(x)\equiv 1-2b\cdot x+b^2x^2$. If $\si_{b}(x_i)=0$ for some $i$ then the image of $x_i$ under $C_{b}$ is on $U_\infty$. Therefore as $b$ changes $\si_{b}(x_i)$ passing through $0$ corresponds to $x_i$ `crossing infinity'. For a fixed configuration $\boldsymbol{w}\in V_f$ and $b\in\mathbb{R}^{1,3}$ such that $C_b\boldsymbol{w}\in V_f$ there are up to $2^4=16$ possibilities\footnote{Note that it is not always 16, e.g.\ if the configuration contains $x=0$ then $\si_b(x)=1$ for all $b$.} for 
\begin{align}\label{3f}
\text{sgn}(\si_b(\boldsymbol{w}))\equiv\text{sgn}(\sp\si_b(x_1),\spa\si_b(x_2),\spa\si_b(x_3),\spa\si_b(x_4)\sp),
\end{align}
and up to eight possibilities for the $6-$tuple $\text{sgn}(k_b(\boldsymbol{w}))$, where $k_b(\boldsymbol{w})\equiv$
\begin{align}\label{2f}
(\si_b(x_1)\si_b(x_2), \si_b(x_3)\si_b(x_4),\si_b(x_2)\si_b(x_3),\si_b(x_1)\si_b(x_4),\si_b(x_1)\si_b(x_3),\si_b(x_2)\si_b(x_4)).
\end{align}
These eight possibilities can be deduced easily by calculating $\text{sgn}(k_b(\boldsymbol{w}))$ for each of the sixteen possibilities for (\ref{3f}).  Letting $S\equiv(+++++\hspace{0.08cm}+)$, these eight possibilities are
\begin{align}\label{4f}
    S,\spac P_{12}P_{14}P_{13}S,\spac P_{12}P_{23}P_{24}S,\spac P_{34}P_{23}P_{13}S,\spac P_{34}P_{14}P_{24}S, 
\end{align}
\begin{align*}
    P_{23}P_{14}P_{13}P_{24}S,\spac     P_{12}P_{34}P_{23}P_{14}S,\spac    P_{12}P_{34}P_{13}P_{24}S.
\end{align*}
For example, if $\text{sgn}(\si_b(\boldsymbol{w}))=(+-+\hspace{0.08cm}-)$ or $(-+-\hspace{0.08cm}+)$ then $\text{sgn}(k_b(\boldsymbol{w}))=(----+\hspace{0.08cm}+)=P_{12}P_{34}P_{23}P_{14}S$. Denoting these eight elements in the order they appear in (\ref{4f}) by $g_i$ for $i=0,1,\dots, 7$, collecting them in a set $G=\{g_i\}_{i=0}^7$, and introducing the composition law\footnote{For example, $(++---\hspace{0.08cm}-)\cdot (+-+-+\hspace{0.08cm}-)=(+--+-\hspace{0.08cm}+)$.}
\begin{align}
 \small   \text{sgn}(a_1, a_2, a_3, a_4, a_5, a_6)\cdot  \text{sgn}(b_1, b_2, b_3, b_4, b_5,b_6)\equiv \text{sgn}(a_1b_1, a_2b_2, a_3b_3, a_4b_4, a_5b_5, a_6b_6)
\end{align}
\normalsize
then $(G,\cdot)$ is an abelian group of order eight, with identity element $g_0=S$. Note that $g_i^2=g_0$ for all $i=0,\dots,7$, and so we conclude that $G\simeq \mathbb{Z}_2\times \mathbb{Z}_2\times \mathbb{Z}_2$. This composition law is compatible with the action of SCTs on any initial kinematics, in the sense that
\begin{align*}
    \text{sgn}(k(C_b\boldsymbol{w}))=  \text{sgn}(k_b(\boldsymbol{w})) \cdot \text{sgn}(k(\boldsymbol{w})),
\end{align*}
for $b\in\mathbb{R}^{1,3}$ and $\boldsymbol{w},C_b\boldsymbol{w}\in V_f$. Therefore the group $G$ encodes the action of SCTs $C_b$ on $\text{sgn}(k(\boldsymbol{w}))$. We define the eight sets
\begin{align}
    K_1\equiv G,\spac K_2\equiv P_{12}G,\spac K_3\equiv P_{23}G,\spac  K_4\equiv P_{13}G,
\end{align}
\begin{align*}
    \bar{K}_i\equiv PK_i,\spac i=1,2,3,4,
\end{align*}
where $P_{12}G\equiv\{P_{12}g_i\}_{i=0}^7$, and similarly for the others. Each of these sets has eight elements, and together they account for the $8\times 8=64$ different possibilities for $\text{sgn}(k(\boldsymbol{w}))$ for a finite configuration $\boldsymbol{w}$. They are listed explicitly in appendix \ref{kin}. Furthermore they are each invariant under the action\footnote{The action of $G$ on these sets is defined analogously to the action of $G$ on itself.} of $G$, and so the eight sets of signs of kinematics $K_i, \bar{K}_i$ are invariant under SCTs\footnote{A similar argument holds for inversions.}. Therefore given a configuration $\boldsymbol{w}\in V_f$ such that $\text{sgn}(k(\boldsymbol{w}))\in K_i$ (or $\bar{K}_i$) for some $i$ then for any $b\in \mathbb{R}^{1,3}$ such that $C_b\boldsymbol{w}\in V_f$ we have that $\text{sgn}(k(C_b\boldsymbol{w}))\in K_i$ (or $\bar{K}_i$) for the same $i$. It is possible to assign a set $K_i$ (or $\bar{K}_i$) to non-finite configurations. For configurations $\boldsymbol{U}\in V$ which are not finite it is always possible to make an infinitesimal SCT $C_b$ such that $C_b\boldsymbol{U}$ is finite. Then sgn$(k(C_b\boldsymbol{U}))\in K_i$ (or $\bar{K}_i$) is well-defined, and we set sgn($k(\boldsymbol{U}))\in K_i$ (or $\bar{K}_i$) for the same $i$, even though $k(\boldsymbol{U})$ is not well-defined. We define the refined subsets of $V_{z,\bar{z}}$
\begin{align}
    V_{i,z,\bar{z}}\equiv\{\boldsymbol{U}\in V_{z,\bar{z}}\spa | \spa \text{sgn}(k(\boldsymbol{U}))\in K_i\},\spac \bar{V}_{i,z,\bar{z}}\equiv\{\boldsymbol{U}\in V_{z,\bar{z}}\spa | \spa \text{sgn}(k(\boldsymbol{U}))\in \bar{K}_i\},
\end{align}
which are well-defined and each separately invariant under the conformal group by the above discussion. Investigating the sets $V_{i,z,\bar{z}}$ and $\bar{V}_{i,z,\bar{z}}$ in detail we find that there are different constraints on $z$ and $\bar{z}$ for each $i$. The first important fact is that a configuration $\boldsymbol{U}\in V$ can only have $z,\bar{z}\in\mathbb{C}\setminus\mathbb{R}$ if $\boldsymbol{U}\in \bar{V}_{1,z,\bar{z}}$, and so
\begin{align}\label{6h}
V_{\mathbb{C}}=\bigcup_{\substack{z\in \mathbb{H}\\ \bar{z}=z^*}}\bar{V}_{1,z,\bar{z}}.
\end{align}
Note that $\bar{K}_1$ includes the Euclidean sign assignment $PS=(-----\hspace{0.12cm}-)$, and the seven other possible signs of kinematics are found by acting with $G$ on $PS$. (\ref{6h}) is proven using the analogue of conformal planes in Minkowski space in appendix \ref{confplanepseudoeuc}. Furthermore given $\boldsymbol{U}\in V_{i,z,\bar{z}}$ (or $\bar{V}_{i,z,\bar{z}}$) with $z,\bar{z}\in \mathbb{R}$ then $z$ and $\bar{z}$ are constrained to different intervals of $\mathbb{R}$ depending on $i$ (table \ref{table1}). For example, let $\boldsymbol{U}\in V_{4,z,\bar{z}}$ so that sgn($k(\boldsymbol{U}))\in K_4$. Going through all the possibilities for $\text{sgn}(k(\boldsymbol{U}))$ in table \ref{table7} we deduce that $u(\boldsymbol{U})<0$ and $v(\boldsymbol{U})<0$. Since $u=z\bar{z},\sp v=(1-z)(1-\bar{z}),$ and we take $\bar{z}\geq z$ we conclude that $z\in(-\infty,0),\sp \bar{z}\in(1,\infty)$. Conversely given a configuration $\boldsymbol{U}\in V$ with a known $z$ and $\bar{z}$ we immediately have information about the signs of the kinematics of $\boldsymbol{U}$. For example, if we are told $\boldsymbol{U}$ has $z=-1,\bar{z}=2$, we see from table \ref{table1} that there are two possibilities $\boldsymbol{U}\in V_{4,-1,2}$ or $\bar{V}_{4,-1,2}$, and so $\text{sgn}(k(\boldsymbol{U}))\in K_{4}$ or $\bar{K}_4$.
\setlength{\extrarowheight}{1.5pt}
\begin{table}
\begin{center}
 \begin{tabular}{|c|c|} 
 \hline
   $i$ & $z,\bar{z}$ for non-empty $V_{i,z,\bar{z}},\sp\sp\bar{V}_{i,z,\bar{z}}$\\ [0.2ex] 
 \hline
  $1$ & $z,\bar{z}\in (-\infty,0)\spa\text{OR}\spa (0,1)\spa\text{OR}\spa (1,\infty)$\\
 \hline
  $2$ & $z\in(-\infty,0),\bar{z}\in (0,1)$\\
 \hline 
 $3$ & $z\in(0,1),\bar{z}\in (1,\infty)$\\
 \hline
 $4$ & $z\in(-\infty,0),\bar{z}\in (1,\infty)$\\
 \hline
\end{tabular}
\caption{$z,\bar{z}$ such that $V_{i,z,\bar{z}}$ is non-empty for $z,\bar{z}\in\mathbb{R}$. Note for $i=1$ we always take $z\leq \bar{z}$.}
\label{table1}
\end{center}
\end{table}
Therefore we define the sets
\begin{align}
     V_1=\bigcup_{\substack{z,\bar{z}\in (-\infty,0)\\z,\bar{z}\in (0,1)\\z,\bar{z}\in (1,\infty)\\z\leq\bar{z}}}V_{1,z,\bar{z}}, \spac \bar{V}_1=\bigcup_{\substack{z,\bar{z}\in (-\infty,0)\\z,\bar{z}\in (0,1)\\z,\bar{z}\in (1,\infty)\\z\leq\bar{z}}}\bar{V}_{1,z,\bar{z}}, 
\end{align}
\begin{align}
     V_2=\bigcup_{\substack{z\in (-\infty,0)\\\bar{z}\in (0,1)}}V_{2,z,\bar{z}}, \spac \bar{V}_2=\bigcup_{\substack{z\in (-\infty,0)\\\bar{z}\in (0,1)}}\bar{V}_{2,z,\bar{z}},
\end{align}
\begin{align}
     V_3=\bigcup_{\substack{z\in (0,1)\\\bar{z}\in (1,\infty)}}V_{3,z,\bar{z}}, \spac \bar{V}_3=\bigcup_{\substack{z\in (0,1)\\\bar{z}\in (1,\infty)}}\bar{V}_{3,z,\bar{z}},
\end{align}
\begin{align}
     V_4=\bigcup_{\substack{z\in (-\infty,0)\\\bar{z}\in (1,\infty)}}V_{4,z,\bar{z}}, \spac \bar{V}_4=\bigcup_{\substack{z\in (-\infty,0)\\\bar{z}\in (1,\infty)}}\bar{V}_{4,z,\bar{z}},
\end{align}
and deduce the decomposition\footnote{A similar decomposition appeared recently in \cite{qiao2020classification} by Jiaxin Qiao while this was being written.}
\begin{align}\label{7h}
    V=\bigcup_{i=1}^4 (V_i\cup\bar{V}_i)\cup V_{\mathbb{C}}.
\end{align}
The question remains: Does $\text{Conf}(\mathbb{R}^{1,3})$ act transitively on each non-empty $V_{i,z,\bar{z}}$ (and $\bar{V}_{i,z,\bar{z}}$)? The answer is yes \textit{up to the discrete transformations $\mathcal{P},\sp\mathcal{T}$}. The proof relies on Minkowskian conformal planes.
\newline
\newline
We define the five \textit{Minkowskian conformal plane} configurations $\boldsymbol{w}_{\mathbb{C}}(r,\phi)$ and $\boldsymbol{w}_{bc}(a,\eta)$ for $b,c\in\{+,-\}$ in table \ref{table2}. They are expressed in terms of the unit vectors $e_0=(1,0,0,0)$ and $e_3=(0,0,0,1)$.
\begin{table}[h!]
\begin{center}
 \begin{tabular}{|c|c|c|} 
 \hline
   Configuration & $(x_1,x_2,x_3,x_4)$ & $z,\bar{z}$\\ [0.2ex] 
 \hline
 \hline
  $\boldsymbol{w}_{\mathbb{C}}(r,\phi)$ & $(0,\sp\sp r(0,\sin\phi,0,\cos\phi),\sp\sp e_3,\sp\sp \iota)$ & $re^{i\phi},re^{-i\phi}$\\
 \hline
  $\boldsymbol{w}_{++}(a,\eta)$ & $(0,\sp\sp a(\cosh\eta,0,0,\sinh\eta),\sp\sp e_0,\sp\sp \iota)$ &$
  ae^{-\eta},ae^{\eta}$\\
 \hline 
 $\boldsymbol{w}_{--}(a,\eta)$ & $(0,\sp\sp a(\sinh\eta,0,0,\cosh\eta),\sp\sp e_3,\sp\sp \iota)$ &$ae^{-\eta},ae^{\eta}$\\
 \hline
 $\boldsymbol{w}_{-+}(a,\eta)$ & $(0,\sp\sp a(\sinh\eta,0,0,\cosh\eta),\sp\sp e_0,\sp\sp \iota)$ &$-ae^{-\eta},ae^{\eta}$\\
 \hline
 $\boldsymbol{w}_{+-}(a,\eta)$ & $(0,\sp\sp a(\cosh\eta,0,0,\sinh\eta),\sp\sp e_3,\sp\sp \iota)$ &$-ae^{-\eta},ae^{\eta}$\\
 \hline
\end{tabular}
\caption{Minkowskian conformal plane configurations.}
\label{table2}
\end{center}
\end{table}
Similarly to the Euclidean case we always find $A\in \text{Conf}(\mathbb{R}^{1,3})$ and possibly a discrete transformation $\mathcal{P},\sp\mathcal{T},$ or $\mathcal{P}\mathcal{T}$ to map configurations in the nine sets in (\ref{7h}) to one of the five configurations in table \ref{table2}. The case of $V_{\mathbb{C}}$ is totally analogous to the Euclidean case, and indeed any configuration $\boldsymbol{U}\in V_{\mathbb{C}}$ with $z=re^{i\phi}, \bar{z}=re^{-i\phi}$ can be conformally mapped (no need for discrete transformations) to the pseudo-Euclidean configuration $\boldsymbol{w}_{\mathbb{C}}(r,\phi)\in\bar{V}_{1,z,\bar{z}}$ for $r>0, \phi\in(0,\pi)$, see appendix \ref{confplanepseudoeuc} for some details. The eight remaining sets in (\ref{7h}) can be conformally mapped to the four sets $\boldsymbol{w}_{bc}(a,\eta)$. This is a bit more subtle, and care must be taken to restrict the range of $a$ and $\eta$ in such a way that each $\boldsymbol{U}\in V_{i,z,\bar{z}}$ (or $\bar{V}_{i,z,\bar{z}}$) is mapped to exactly one configuration $\boldsymbol{w}_{bc}(a,\eta)$ for the appropriate $b,c$. This can be done provided the discrete transformations $\mathcal{P},\sp\mathcal{T}$ are used. The results are summarised in table \ref{table3}.
\begin{table}[h!]
\begin{center}
 \begin{tabular}{|c|c|c|c|} 
 \hline
   $V$ & Mapped to &  Configuration bounds & $z,\bar{z}$ bounds\\ [0.2ex] 
 \hline
  $V_{\mathbb{C}}$ & $\boldsymbol{w}_{\mathbb{C}}(r,\phi)$ & $r>0, \phi\in(0,\pi)$ & $z\in \mathbb{H}$, $\bar{z}=z^*$\\
 \hline 
  &  & $a\in(0,1),\sp e^\eta\in(1,1/a)$ & $0<z\leq\bar{z}<1$ \\
  $V_1$ & $\boldsymbol{w}_{++}(a,\eta)$ &$a\in(1,\infty),\sp e^\eta\in(1,a)$ & $1<z\leq\bar{z}<\infty$ \\
  & & $a\in(-\infty,0),\sp e^\eta\in(1,\infty)$ & $-\infty<z\leq\bar{z}<0$\\
  \hline
  $V_{2}$ & $\boldsymbol{w}_{-+}(a,\eta)$ & $a\in(0,\infty),e^{\eta}\in(0,1/a)$&$-\infty<z<0<\bar{z}<1$\\
 \hline
 $V_{3}$ & $\boldsymbol{w}_{--}(a,\eta)$ &$a\in(0,\infty),e^\eta\in(\text{max}(a,1/a),\infty)$& $0<z<1<\bar{z}<\infty$ \\
 \hline
 \hline
  $V_{4}$ & $\boldsymbol{w}_{+-}(a,\eta)$ & $a\in(0,\infty),e^{\eta}\in(1/a,\infty)$&$-\infty<z<0<1<\bar{z}<\infty$\\
 \hline
\end{tabular}
\caption{Conformal plane structure for $V_{\mathbb{C}}$ and $V_i$. For $\bar{V}_i$ the only change is that the signs on $\boldsymbol{w}_{bc}(a,\eta)$ reverse. For example, $\bar{V}_1$ gets mapped to $\boldsymbol{w}_{--}(a,\eta)$, and all other columns with respect to $V_1$ are unchanged.}
\label{table3}
\end{center}
\end{table}
\newline
We explain the argument for $V_2$, so let $\boldsymbol{w}=(x_1,x_2,x_3,x_4)\in V_2$. Without loss of generality we can take $\boldsymbol{w}$ to be a finite configuration. If $\boldsymbol{w}$ is not a finite configuration we can first make a small SCT such that it is finite. As in the Euclidean case we define the conformal transformation\footnote{We can make $A_1$ explicitly connected to the identity by making an extra inversion and using $C_b=\mathcal{I}T_{-b}\mathcal{I}$} $A_1\equiv T_{-\mathcal{I}x_1}\mathcal{I}T_{-x_4}$ and act on $\boldsymbol{w}$
\begin{align}
   \boldsymbol{w}\rightarrow \boldsymbol{w}_1\equiv A_1\boldsymbol{w}=(0,y_2,y_3,\iota),
\end{align}
where $y_2\equiv A_1x_2=\mathcal{I}(x_{24})-\mathcal{I}(x_{14}),\sp\sp y_3\equiv A_1x_3=\mathcal{I}(x_{34})-\mathcal{I}(x_{14})$. The stabiliser of $0$ and $\iota$ is generated by dilatations and Lorentz rotations $SO^+(1,3)\subset SO^+(2,4)$. How we proceed next depends on the signs of the remaining free kinematics. We compute 
\begin{align}
    (y_{2}^2,y_{3}^2,y_{23}^2)=\left(\frac{x_{12}^2}{x_{14}^2x_{24}^2},\frac{x_{13}^2}{x_{14}^2x_{34}^2},\frac{x_{23}^2}{x_{24}^2x_{34}^2}\right),
\end{align}
where we used the property that under inversions the kinematics transform
\begin{align}
    \mathcal{I}:x_{ij}^2\longrightarrow \frac{x_{ij}^2}{x_i^2x_j^2}.
\end{align}
Since $\boldsymbol{w}\in V_2$ we have information about the signs of the kinematics, namely that $\text{sgn}(k(\boldsymbol{w}))\in K_2$ (see table \ref{table7}). For example, we could have $\text{sgn}(k(\boldsymbol{w}))=P_{12}S=(-++++\sp\sp+)$. In this case we have that 
\begin{align}\label{8h}
    \text{sgn}(y_{2}^2,y_{3}^2,y_{23}^2)=(-++).
\end{align}
In fact going through all the cases in table \ref{table7} it turns out that (\ref{8h}) holds for any $\text{sgn}(k(\boldsymbol{w}))\in K_2$, and hence $\boldsymbol{w}\in V_2$. We have that $y_3^2=c^2>0$, so we can use a Lorentz rotation $L$ to map $y_3\rightarrow (\pm c,0,0,0)$ and then rescale to $\pm e_0$. Defining $A_2\equiv D_{1/c}L$ we map
\begin{align}
    \boldsymbol{w}_1\rightarrow \boldsymbol{w}_2^\pm\equiv A_2\boldsymbol{w}_1=(0,\sp\sp z_2,\sp\sp\pm e_0,\sp\sp\iota),
\end{align}
where $z_2\equiv A_2y_2$. The stabiliser of $0, \pm e_0,$ and $\iota$ is generated by rotations $SO(3)$ acting on the Euclidean coordinates of $\mathbb{R}^{1,3}$. We use an $SO(3)$ rotation $R$ to eliminate the second and third component of $z_2$, so $z_2\rightarrow (z_2^0,0,0,z_2^3)$. We know from (\ref{8h}) that $(z_2^0)^2-(z_2^3)^2<0$, so we map
\begin{align}\label{9h}
\boldsymbol{w}_2^{\pm}\rightarrow \bar{\boldsymbol{w}}^{\pm}(a,\eta)\equiv R\boldsymbol{w}_{2}^\pm=(0,\sp\sp a(\sinh\eta,\sp0,\sp0,\sp\cosh\eta),\sp\sp\pm e_0,\sp\sp\iota),     
\end{align}
for some $a\in \mathbb{R}\setminus \{0\}, \eta\in \mathbb{R}$. We compute the conformal invariants of $
\bar{\boldsymbol{w}}^{\pm}(a,\eta)$ to be
\begin{align}\label{10h}
    z=\text{min}(-ae^{\mp\eta},ae^{\pm\eta}),\spac \bar{z}=\text{max}(-ae^{\mp\eta},ae^{\pm\eta}).
\end{align}
Note we have the issue that starting with a configuration $\boldsymbol{w}\in V_2$ we have two possible configurations (\ref{9h}) we can end up with, $\bar{\boldsymbol{w}}^{+}(a,\eta)$ and $\bar{\boldsymbol{w}}^-(a,\eta)$. Moreover we know from table \ref{table1} that $z\in (-\infty,0), \bar{z}\in (0,1)$. We need our final configuration $\boldsymbol{w}_{-+}(a,\eta)$ to contain each possible set of conformal invariants exactly once. This is because we want to be sure we are mapping all $\boldsymbol{w}\in V_{2,z,\bar{z}}$ to the exact same configuration, to prove that $\text{Conf}(\mathbb{R}^{1,3})$ acts transitively on $V_{2,z,\bar{z}}$. A priori there could be multiple configurations with the same $z,\bar{z}$ in our final $\boldsymbol{w}_{-+}(a,\eta)$. To resolve these issues we need to use the discrete transformations $\mathcal{P}$ and $\mathcal{T}$. We first note that $\bar{\boldsymbol{w}}^+(a,\eta)=\mathcal{T}\bar{\boldsymbol{w}}^-(a,-\eta)$. Therefore if we end up at $\bar{\boldsymbol{w}}^-(a,\eta)$ we simply apply a time-reversal transformation and relabel $\eta\rightarrow \eta'\equiv -\eta$ to arrive at $\bar{\boldsymbol{w}}^+(a,\eta)$. We can invert (\ref{10h}) to recover $a$ and $\eta$ for $\bar{\boldsymbol{w}}^+(a,\eta)$
\begin{align}{\label{11h}}
    a=\pm\sqrt{-z\bar{z}},\spac \eta=\pm\frac{1}{2}\log(-\bar{z}/z).
\end{align}
We see from (\ref{11h}) that indeed multiple configurations $\bar{\boldsymbol{w}}^+(a,\eta)$ give the same $z,\bar{z}$, which we want to avoid. We note that $\bar{\boldsymbol{w}}^+(a,\eta)=\mathcal{P}\bar{\boldsymbol{w}}^+(-a,-\eta)$. Therefore any configuration with $a<0$ we can apply a parity transformation and relabel $\eta\rightarrow \eta'=-\eta$ at no cost, so we can assume $a>0$. In this case we have that
\begin{align}\label{12h}
    z=-ae^{-\eta},\spac \bar{z}=ae^{\eta},\spac a=\sqrt{-z\bar{z}},\spac \eta=\frac{1}{2}\log(-\bar{z}/z).
\end{align}
From (\ref{12h}) we see that we have exactly one configuration $\bar{\boldsymbol{w}}^{+}(a>0,\eta)\equiv \boldsymbol{w}_{-+}(a,\eta)$ for each $z,\bar{z}$, as required. To get $z,\bar{z}$ in the correct range $-\infty<z<0<\bar{z}<1$, or equivalently enforce the $y_{23}^2>0$ condition of (\ref{8h}), there is a further constraint $e^{\eta}<1/a$. The final configuration $\boldsymbol{w}_{-+}(a,\eta)$ in terms of $z,\bar{z}$ is
\begin{align}
    \boldsymbol{w}_{-+}(z,\bar{z})=\left(0,\sp\sp \left(\frac{\bar{z}+z}{2},0,0,\frac{\bar{z}-z}{2}\right),\sp\sp e_0,\sp\sp \iota\right),
\end{align}
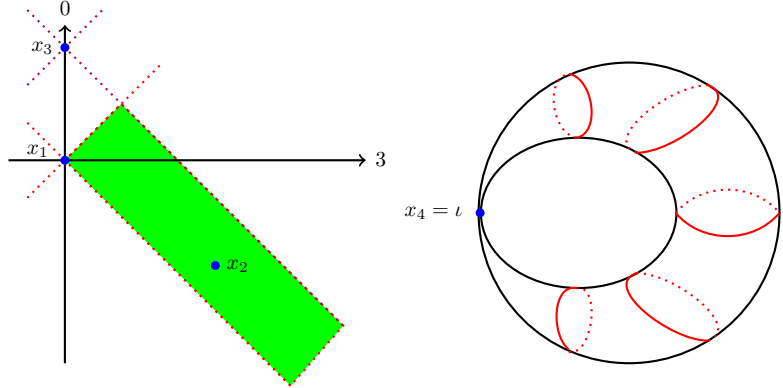
\begin{figure}[h!]
\begin{center}
\begin{tikzpicture}[thick,scale=1, every node/.style={scale=0.81}]
\draw  [red,dotted,fill=green] (-5.5,0.7)-- (-4.75,1.45) -- (-1.8,-1.5) -- (-2.5,-2.3) --cycle ;
\draw[->] (-6.25,0.7)--(-1.5,0.7); 
\draw[->] (-5.5,-2)--(-5.5,2.5); 
  \draw[thick] (1.33,0) ellipse (1.3 and 1);
\draw[dotted,red] (-6,1.2)--(-2.5,-2.3);
\draw[dotted,red] (-6,0.2)--(-4.2,2);
\draw[dotted,purple] (-6,2.7)--(-1.8,-1.5);
\draw[dotted,purple]  (-6,1.7)--(-5,2.7);

\filldraw[blue] (-5.5,2.2) circle (1.3pt);
\filldraw[blue] (-5.5,0.7) circle (1.3pt);
\filldraw[blue] (-3.5,-0.7) circle (1.3pt);
\draw[thick] (0,0) arc (180:-180:2);

\draw[red,thick]    (2.62,0) to[out=-50,in=-130] (4,0);
\draw[red,dotted]    (2.62,0) to[out=50,in=130] (4,0);

\draw[red,thick]    (2.1,0.8) to[out=0,in=-20] (3.05,1.7);
\draw[red,dotted]    (3.05,1.7) to[out=180,in=160] (2.1,0.8);
\draw[red,dotted]    (2.1,-0.8) to[out=0,in=20] (3.05,-1.7);
\draw[red,thick]    (3.05,-1.7) to[out=180,in=-160] (2.1,-0.8);

\draw[red,thick]    (1.3,1) to[out=10,in=-10] (1.2,1.85);
\draw[red,dotted]    (1.3,1) to[out=170,in=190] (1.2,1.85);

\draw[red,dotted]    (1.3,-1) to[out=10,in=-10] (1.2,-1.85);
\draw[red,thick]    (1.3,-1) to[out=170,in=140] (1.2,-1.82);

\filldraw[blue] (0.02,0) circle (1.3pt);
\footnotesize
\node at (-0.6,0){$x_4=\iota$};
\node at (-5.8,2.2){$x_3$};
\node at (-5.86,0.83){$x_1$};
\node at (-3.2,-0.7){$x_2$};
\node at (-1.3,0.715){$3$};
\node at (-5.495,2.72){$0$};

\normalsize
\end{tikzpicture}
\end{center}
    \caption{Minkowskian conformal plane configuration $\boldsymbol{w}_{-+}(a,\eta)$ for $V_2$, $z\in (-\infty,0),\sp\bar{z}\in(0,1)$. $x_2$ can be anywhere in green region depending on $z,\bar{z}$.}
    \label{figg22}
\end{figure}\noindent and is shown in figure \ref{figg22}. 
\newline
These arguments can be repeated analogously for the other sets $V_{i,z,\bar{z}}$ and $\bar{V}_{i,z,\bar{z}}$ to show that up to the discrete transformations $\mathcal{P},\mathcal{T}$ all configurations in these sets are conformally equivalent to a single Minkowskian conformal plane configuration $\boldsymbol{w}_{b,c}(a,\eta)$ or $\boldsymbol{w}_\mathbb{C}(r,\phi)$, as summarised in table \ref{table3}. Note that for $\bar{V}_{1,z,\bar{z}}$ there are two cases, one where $z,\bar{z}\in\mathbb{R}$ and one where $z,\bar{z}\in\mathbb{C}\setminus \mathbb{R}$. Therefore up to the discrete transformations $\mathcal{P},\mathcal{T}$ $\text{Conf}(\mathbb{R}^{1,3})$ acts transitively on each non-empty $V_{i,z,\bar{z}}$ and $\bar{V}_{i,z,\bar{z}}$. Given configurations $\boldsymbol{U}_1,\boldsymbol{U}_2\in V_{i,z,\bar{z}}$ (or $\bar{V}_{i,z,\bar{z}}$) we deduce from the above the existence of $A_i\in \text{Conf}(\mathbb{R}^{1,3})$ and discrete transformations $\mathcal{D}_i\in\{\mathbb{I},\mathcal{P},\mathcal{T},\mathcal{P}\mathcal{T}\}$ for $i=1,2$ such that $\mathcal{D}_1A_1\boldsymbol{U}_1=\mathcal{D}_2A_2\boldsymbol{U}_2=\bar{\boldsymbol{w}}(z,\bar{z})$ for exactly one Minkowskian conformal plane configuration $\bar{\boldsymbol{w}}(z,\bar{z})$ given in table \ref{table2}. Then if we let $A\equiv (\mathcal{D}_1A_1)^{-1}\mathcal{D}_2A_2$ we see that $\boldsymbol{U}_1=A\boldsymbol{U}_2$.
\newline
\newline
Overall we have shown that all configurations $\boldsymbol{U}\in V_{z,\bar{z}}$ with $z,\bar{z}\in\mathbb{C}\setminus\mathbb{R}$ are conformally equivalent, and that for $\boldsymbol{U}\in V_{z,\bar{z}}$ with $z,\bar{z}\in\mathbb{R}$ there are two conformal equivalence classes of configurations (up to $\mathcal{P},\sp\mathcal{T}$) exchanged by a reversal of kinematics $V_{i,z,\bar{z}}\rightarrow \bar{V}_{i,z,\bar{z}}=PV_{i,z,\bar{z}}$ for a fixed $i$.

\section{Symmetries of the Box Integral}\label{symmetry}
Let $\boldsymbol{w}=(x_1,x_2,x_3,x_4)\in V_f$ be a finite configuration. The dual conformal box integral in Minkowski space is defined
\begin{align}\label{5b}
   I(\boldsymbol{w})=\int\frac{d^4x_5}{i\pi^2}\frac{x_{13}^2x_{24}^2}{(x_{15}^2+ i\epsilon)(x_{25}^2+ i\epsilon)(x_{35}^2 + i\epsilon)(x_{45}^2+ i\epsilon)},
\end{align}
where $+i\epsilon$ indicates the Feynman prescription for propagators in Minkowski space ($\epsilon$ is always taken to be a positive infinitesimal quantity). The chosen prescription tells us on which side of the branch cut we must evaluate any multivalued functions that appear upon computing the integral. It is important to take $\boldsymbol{w}$ to be a finite configuration because the integral is not well-defined when any of the points\footnote{The only exception is the case where a single point is at $\iota$. In this case the integral is well-defined.} $x_i$ are on $U_{\infty}$. In the next section we will consider specific lightlike limits of (\ref{5b}) such that two of the external points are sent to $\mathscr{I}$. However the result depends in general on the direction these points are sent to $\mathscr{I}$, so that it is not possible to define the integral if one or more points are exactly on $\mathscr{I}$. Configurations with one or more points on $\mathscr{I}$ are singular and are exactly the configurations where the value of (\ref{5b}) can change under infinitesimal SCTs.
\newline
\newline
We study the symmetries of the integral (\ref{5b}) directly in Minkowski space. There are two main different representations useful for studying these symmetries, given in \cite{Duplan_i__2002}. The first is the Feynman parametric representation
\begin{align}\label{9b}
    I(\boldsymbol{w})= I'(x_{ij}^2)\equiv  x_{13}^2x_{24}^2\int_0^1 d\boldsymbol{y}\frac{\delta(1-y_1-y_2-y_3-y_4)}{(\sum_{j<k}y_jy_kx_{jk}^2+ i\epsilon)^2},
\end{align}
where $\int_0^{1}d\boldsymbol{y}\equiv \prod_{i=1}^4\int_0^1 dy_i$. Note that $I'$ is manifestly a function of the kinematics $x_{ij}^2$. We also see that
\begin{align}\label{8b}
    I'(-x_{ij}^2)=I'(x_{ij}^2)^*.
\end{align}
Another useful expression for (\ref{5b}) is the Mellin representation $I'(x_{ij}^2)=$
\begin{align}\label{10b}
   \frac{1}{(2\pi i)^2}\int_{C_1}ds\int_{C_{2}}ds'(\Gamma_{s,s'})^2 \left(\frac{x_{12}^2\!+\!i\epsilon}{x_{13}^2\!+\!i\epsilon}\right)^{-s}\left(\frac{x_{34}^2\!+\!i\epsilon}{x_{24}^2\!+\!i\epsilon}\right)^{-s}\left(\frac{x_{14}^2\!+\!i\epsilon}{x_{13}^2\!+\!i\epsilon}\right)^{-s'}\left(\frac{x_{23}^2\!+\!i\epsilon}{x_{24}^2\!+\!i\epsilon}\right)^{-s'},
\end{align}
where $\Gamma_{x,y}\equiv \Gamma(x)\Gamma(y)\Gamma(1-x-y)$. The complex power is $z^s\equiv\exp(s\log z)$, where the logarithm has a branch cut on the negative real axis. $C_j$ are contours in the complex plane
\begin{align}
    C_j=\{\ga_j+it\sp|\sp t\in(-\infty,\infty)\},
\end{align}
where $\ga_1,\ga_2$ satisfy $0<\ga_1,\ga_2<1,\sp 0<\ga_1+\ga_2<1$. The corresponding Euclidean versions of (\ref{9b}) and (\ref{10b}) are found by setting $\epsilon=0$.
\subsection{Conformal Transformations}\label{confbox}
$\text{Conf}(\mathbb{R}^{1,3})$ acts on $\mathbb{R}^{1,3}_c$ by translations, proper orthochronous Lorentz rotations, dilatations, and SCTs. Since (\ref{9b}) and (\ref{10b}) are functions of only the kinematics $x_{ij}^2$, which are Poincaré invariants, it is clear that (\ref{5b}) is invariant under Poincaré transformations
\begin{align}
    I(\boldsymbol{w})=I(T_a\boldsymbol{w})=I(L\boldsymbol{w}),
\end{align}
for all $a\in\mathbb{R}^{1,3},\spa L\in SO^+(1,3)$. It is also clear that (\ref{5b}) is invariant under dilatations $x_i\rightarrow D_\la x_i\equiv \la x_i$, since for all $\la>0$ we have
\begin{align}
 I(D_\la\boldsymbol{w})=\int \frac{d^4x_5}{i\pi^2}\frac{\la^2x_{13}^2\la^2x_{24}^2}{\prod_{j=1}^4((\la x_j-x_5)^2+ i\epsilon)}
\end{align}
\begin{align*}
 =(x_5=\la x_5')=\int \frac{\la^4 d^4x_5'}{i\pi^2\la^8}\frac{\la^4x_{13}^2x_{24}^2}{\prod_{j=1}^4(( x_j-x_5')^2+ i\frac{\epsilon}{\la^2})}
\end{align*}
\begin{align*}
=\int \frac{ d^4x_5'}{i\pi^2}\frac{x_{13}^2x_{24}^2}{\prod_{j=1}^4(( x_j-x_5')^2+ i\epsilon')}=I(\boldsymbol{w}),
\end{align*}
because $\epsilon'\equiv \frac{\epsilon}{\la^2}$ is still an infinitesimal positive quantity. This invariance under dilatations is also easily proven from representations (\ref{9b}) and (\ref{10b}). In Euclidean signature (\ref{3t}) the box integral is also checked to be invariant under SCTs $C_b$, and so the integral is fully conformally invariant. In Minkowski space however this invariance is broken by the IR regulator $i\epsilon$. Exactly how the invariance is broken is most easily seen in terms of the Mellin representation (\ref{10b}). For $b\in \mathbb{R}^{1,3}$ we have that $I'(x_{ij}^2)\rightarrow  I'(C_bx_{ij}^2)=$
\begin{align}\label{13j}
   \frac{1}{(2\pi i)^2}\int_{C_1}ds\int_{C_{2}}ds'(\Gamma_{s,s'})^2 \left(\frac{x_{12}^2\!+\!i\epsilon_{12}}{x_{13}^2\!+\!i\epsilon_{13}}\right)^{-s}\left(\frac{x_{34}^2\!+\!i\epsilon_{34}}{x_{24}^2\!+\!i\epsilon_{24}}\right)^{-s}\left(\frac{x_{14}^2\!+\!i\epsilon_{14}}{x_{13}^2\!+\!i\epsilon_{13}}\right)^{-s'}\left(\frac{x_{23}^2\!+\!i\epsilon_{23}}{x_{24}^2\!+\!i\epsilon_{24}}\right)^{-s'},
\end{align} 
\begin{align}
    \epsilon_{ij}\equiv \epsilon\si_b(x_i)\si_b(x_j),
\end{align}
where (\ref{10f}) was used and $\si_b(x)=1-2b\cdot x+b^2x^2$. From (\ref{13j}) we see the extent to which conformal invariance is broken is encoded in the signs of the infinitesimal imaginary parts $i\epsilon_{ij}$. Given a finite configuration $\boldsymbol{w}=(x_1,x_2,x_3,x_4)$ there is a possible branch jumping of the integral under \textit{finite} special conformal transformations $C_b$ for which at least one of the $\epsilon_{ij}<0$. $\epsilon_{ij}$ changes from positive to negative when $C_bx_k$ crosses infinity for some $k\in\{i,j\}$ and the kinematics $x_{kl}^2$ for $l\neq k$ change sign, i.e.\ the kinematic region changes. In computing (\ref{13j}) the $i\epsilon_{ij}$ select the correct branch of the logs/dilogs which appear in the computation, after which the result depends only on the cross-ratios $u$ and $v$, or equivalently $z$ and $\bar{z}$. This computation was completed for arbitrary kinematics in \cite{Duplan_i__2002}, and we specialise it in section \ref{branches} to give the box integral as a function of $z$ and $\bar{z}$ in each kinematic region $\text{sgn}(k(\boldsymbol{w}))$. While from (\ref{13j}) there are naively eight values of the integral which can be reached under SCTs, corresponding to the eight possible sign assignments for $\epsilon_{ij}$, it is actually up to four. The dependence of the box integral on only the conformal invariants $z$ and $\bar{z}$ as well as the kinematic region $\text{sgn}(k(\boldsymbol{w}))$ can be thought of as \textit{pseudo}-conformal invariance.
\newline
\newline
As for discrete conformal transformations, the parity and time reversal maps $\mathcal{P},\mathcal{T}:\mathbb{R}^{1,3}\rightarrow \mathbb{R}^{1,3}$ are also symmetries of the box integral. $\mathcal{P}$ and $\mathcal{T}$ do not change the kinematics $x_{ij}^2$, so
\begin{align}
  I(\boldsymbol{w})=I(\mathcal{P}\boldsymbol{w})=I(\mathcal{T}\boldsymbol{w})=I(\mathcal{P}\mathcal{T}\boldsymbol{w}).
\end{align}
It is also useful to keep in mind how $z,\bar{z}$, and $I(\boldsymbol{w})$ respond to permutations $\sigma\in S_4$ of the external points. This is discussed in appendix \ref{permutation}. 

\subsection{Branches of the Minkowski Box Integral}\label{branches}
We write the result of \cite{Duplan_i__2002} for $I(\boldsymbol{w})$ in all kinematic regions explicitly in terms of the conformal invariants $z,\bar{z}$. We first note that given a finite configuration $\boldsymbol{w}\in V_f$ with real $z,\bar{z}$ it is always possible to find a permutation $\si\in S_4$ such that $\si\boldsymbol{w}$ either has $z,\bar{z}\in (0,1)$, or $z\in (-\infty,0),\bar{z}\in(0,1)$. Since $I(\boldsymbol{w})$ transforms at most by a conformally invariant constant\footnote{Since it is conformally invariant, $I(\boldsymbol{w})$ and $I(\si\boldsymbol{w})$ have the same qualitative behaviour under SCTs.} under this permutation (table \ref{table5}), we can focus our attention on \textit{restricted} configurations $\boldsymbol{w}$ with $z,\bar{z}\in\mathbb{C}$, $z,\bar{z}\in (0,1)$, or $z\in (-\infty,0),\bar{z}\in(0,1)$ so that $\boldsymbol{w}$ is either in $V_{\mathbb{C}},V_1,\bar{V}_1, V_2,$ or $\bar{V}_2$ as defined in section \ref{confplanemink}. We define four functions $f_i(z,\bar{z})$ by
\begin{align}
    f_1(z,\bar{z})=1,\spac  f_2(z,\bar{z})=\log(z/\bar{z}),\spac  f_3(z,\bar{z})=\log(\frac{1-z}{1-\bar{z}}),
\end{align}
\begin{align}
    f_4(z,\bar{z})=2\text{Li}_2(z)-2\text{Li}_2(\bar{z})+\log(z\bar{z})\log(\frac{1-z}{1-\bar{z}}),
\end{align}
where we take $z,\bar{z}\in \mathbb{C}\setminus [1,\infty),$ $\bar{z}\geq z$ if $z,\bar{z}\in \mathbb{R}$, and $z\in\mathbb{H},\bar{z}=z^*$ if $z,\bar{z}\in\mathbb{C}\setminus\mathbb{R}$. Note these are the same functions which appear from Yangian invariance in \cite{Loebbert_2020} and $f_1,f_2,f_3$ are essentially the functions arising from period contours in \cite{hodges2010box}. We evaluate arguments on the branch cut of the logarithms (the negative real axis) slightly above the cut, so we implicitly take $\log(z)\rightarrow\log(z+i\epsilon)$. Due to the way we organised things, we never need to consider arguments on the branch cut $(1,\infty)$ of $\text{Li}_2(z)$. Note that $f_4$ is the Bloch-Wigner function (\ref{4t}) when $\bar{z}=z^*$. In this section, given a finite configuration $\boldsymbol{w}$, we abbreviate the kinematic region $k\equiv \text{sgn}(k(\boldsymbol{w}))$. For any restricted configuration we can write the box integral (\ref{5b}) in terms of $f_i(z,\bar{z})$ 
\begin{align}\label{4i}
    I(z,\bar{z},k)=\sum_{i=1}^4\frac{c_i^kf_i(z,\bar{z})}{z-\bar{z}},
\end{align}
where $c_i^k\in\mathbb{C}$ depend on the kinematic region $k$. We specialise the result of \cite{Duplan_i__2002} to restricted configurations, i.e.\ those in $V_{\mathbb{C}}$, $V_1$ with $z,\bar{z}\in (0,1)$, $\bar{V}_1$ with $z,\bar{z}\in(0,1)$, $V_2$, and $\bar{V}_2$.
\begin{table}[h!]
\begin{center}
 \begin{tabular}{||c|c|c|c||} 
 \hline
 $k$ & $I(z,\bar{z},k)$ \\ [0.5ex] 
 \hline
$(+++++\hspace{0.1cm}+),(-++--\hspace{0.1cm}+)$, &\\
$(-+-++\hspace{0.1cm}-)$, &   $\frac{f_4}{z-\bar{z}}$\\ 
$(+--+-\hspace{0.1cm}+),(+-+-+\hspace{0.1cm}-)$&\\
\hline
$(++---\hspace{0.1cm}-)$ &$\frac{f_4-2\pi i f_3}{z-\bar{z}}$\\
\hline
$(--++-\hspace{0.1cm}-)$ & $\frac{f_4+2\pi i f_2}{z-\bar{z}}$\\
\hline
$(----+\hspace{0.1cm}+)$ & Missing region\\
\hline
\end{tabular}
\caption{$I(z,\bar{z},k)$ for $0<z\leq\bar{z}<1$, $k\in K_1$.}
\label{table8}
\end{center}
\end{table}
\begin{table}[h!]
\begin{center}
 \begin{tabular}{||c|c|c|c||} 
 \hline
 $k$ & $I(z,\bar{z},k)$ \\ [0.5ex] 
 \hline
$(-----\hspace{0.1cm}-),(+--++\hspace{0.1cm}-)$, &\\
$(+-+--\hspace{0.1cm}+)$, &   $\frac{f_4}{z-\bar{z}}$\\ 
$(-++-+\hspace{0.1cm}-),(-+-+-\hspace{0.1cm}+)$&\\
\hline
$(--+++\hspace{0.1cm}+)$ &$\frac{f_4+2\pi i f_3}{z-\bar{z}}$\\
\hline
$(++--+\hspace{0.1cm}+)$ & $\frac{f_4-2\pi i f_2}{z-\bar{z}}$\\
\hline
$(++++-\hspace{0.1cm}-)$ & $\frac{f_4+2\pi i(f_2-f_3-2\pi i f_1)}{z-\bar{z}}$\\
\hline
\end{tabular}
\caption{$I(z,\bar{z},k)$ for $0<z\leq\bar{z}<1$ or $z\in\mathbb{H},\bar{z}=z^*$, $k\in \bar{K}_1$.}
\label{table9}
\end{center}
\end{table}
\newline
We take a configuration $\boldsymbol{w}\in V_1$ with $0<z\leq\bar{z}<1$, so that $k\in K_1$ (see table \ref{table7}). Firstly we conjecture that such configurations $\boldsymbol{w}$ cannot have $k=k^*\equiv (----+\hspace{0.05cm}+)$, when there is a priori no reason this is not possible. We demonstrate this fact numerically in appendix \ref{missing}. We call this the `missing' kinematic region. Note it is possible for a configuration $\boldsymbol{w}\in V_1$ to have $k=(----++)$ if $z,\bar{z}\in(-\infty,0)$ or $z,\bar{z}\in (1,\infty)$. It is also possible to pick kinematics $x_{ij}^2$ such that $k=(----+\hspace{0.1cm}+)$ and $z,\bar{z}\in (0,1)$, however we conjecture such kinematics cannot be realised by configurations $\boldsymbol{w}\in V_f$. We list the results for the box integral $I(z,\bar{z})$ as a function of the kinematic region $k\in K_1$ in table \ref{table8}. Since Conf$(\mathbb{R}^{1,3})$ acts transitively\footnote{Up to $\mathcal{P},\sp \mathcal{T}$ which doesn't change $I(\boldsymbol{w})$.} on each $V_{1,z,\bar{z}}$ we see that given any configuration $\boldsymbol{w}\in V_1$ with $z,\bar{z}\in (0,1)$ one can reach \textbf{three} branches of the box integral using SCTs. If $\boldsymbol{w}\in V_{\mathbb{C}}$ or $\bar{V}_1$ with $0<z\leq\bar{z}<1$, then $k\in \bar{K}_1$. In this case the kinematics reversed version of the missing region $Pk^*=(++++-\hspace{0.08cm}-)$ with $z,\bar{z}\in(0,1)$ is realisable, although numerically it is found to be much rarer than the other regions (appendix \ref{missing}). We list the results for the box integral $I(z,\bar{z})$ as a function of the kinematic region $k\in \bar{K}_1$ in table \ref{table9}. In this case starting with any configuration $\boldsymbol{w}\in V_{\mathbb{C}}$ or $\bar{V}_1$ with $0<z\leq\bar{z}<1$ one can reach \textbf{four} branches of the box integral with SCTs. For a fixed $z,\bar{z}\in (0,1)$ there are \textbf{six} possible values of the box integral, corresponding to the four functions in table \ref{table9} and the two differing functions in table \ref{table8}. Note that when $f_1$ appears there is no symmetry under $z\leftrightarrow\bar{z}$ and our ordering of $z,\bar{z}$ is important.
\newline
\begin{table}[h!]
\begin{center}
 \begin{tabular}{||c|c|c|c||} 
 \hline
 $k$ & $I(z,\bar{z},k)$ \\ [0.5ex] 
 \hline

$(-++++\hspace{0.1cm}+),(---+-\hspace{0.1cm}+),(--+-+\hspace{0.1cm}-)$ &   $\frac{f_4}{z-\bar{z}}$\\ 
 \hline
$(+++--\hspace{0.1cm}+),(++-++\hspace{0.1cm}-),(-+---\hspace{0.1cm}-)$ &   $\frac{f_4-2\pi i f_3}{z-\bar{z}}$\\ 
\hline
$(+---+\hspace{0.1cm}+)$ & $\frac{f_4-2\pi i f_2}{z-\bar{z}}$\\
\hline
$(+-++-\hspace{0.1cm}-)$ & $\frac{f_4+2\pi i(f_2-f_3-2\pi i f_1)}{z-\bar{z}}$\\
\hline
\end{tabular}
\caption{$I(z,\bar{z},k)$ for $z\in(-\infty,0),\sp \bar{z}\in(0,1)$, $k\in K_2$.}
\label{table10}
\end{center}
\end{table}

\begin{table}[h!]
\begin{center}
 \begin{tabular}{||c|c|c|c||} 
 \hline
 $k$ & $I(z,\bar{z},k)$ \\ [0.5ex] 
 \hline
$(---++\hspace{0.1cm}-),(--+--\hspace{0.1cm}+),(+-+++\hspace{0.1cm}+)$ &   $\frac{f_4}{z-\bar{z}}$\\ 
 \hline
$(+----\hspace{0.1cm}-),(+++-+\hspace{0.1cm}-),(++-+-\hspace{0.1cm}+)$ &   $\frac{f_4-2\pi i f_3}{z-\bar{z}}$\\ 
\hline
$(-+--+\hspace{0.1cm}+)$ & $\frac{f_4-2\pi i f_2}{z-\bar{z}}$\\
\hline
$(-+++-\hspace{0.1cm}-)$ & $\frac{f_4+2\pi i(f_2-f_3-2\pi i f_1)}{z-\bar{z}}$\\
\hline
\end{tabular}
\caption{$I(z,\bar{z},k)$ for $z\in(-\infty,0),\sp \bar{z}\in(0,1)$, $k\in 
\bar{K}_2$.}
\label{table11}
\end{center}
\end{table}
\noindent We list the corresponding results for configurations $\boldsymbol{w}\in V_f$ with $z\in(-\infty,0), \bar{z}\in(0,1)$, so that $k\in K_2$ or $k\in\bar{K}_2$, in tables \ref{table10} and \ref{table11} respectively. In both cases starting with any finite configuration $\boldsymbol{w}\in V_2$ (or $\bar{V}_2$) one can reach \textbf{four} branches of the box integral using SCTs. Since the functions in tables \ref{table10} and \ref{table11} overlap given any finite configuration with $z\in(-\infty,0), \bar{z}\in(0,1)$ there are \textbf{four} possible values for the box integral.

\section{Double Infinity Configurations}\label{doubleinf}
\begin{figure}[!htb]
    \centering
    \begin{minipage}{.5\textwidth}
        \centering
       \begin{tikzpicture}[thick,scale=0.7, every node/.style={scale=0.8}]

\node[inner sep=0pt] (russell) at (0,0)
    {\includegraphics[width=.8\textwidth]{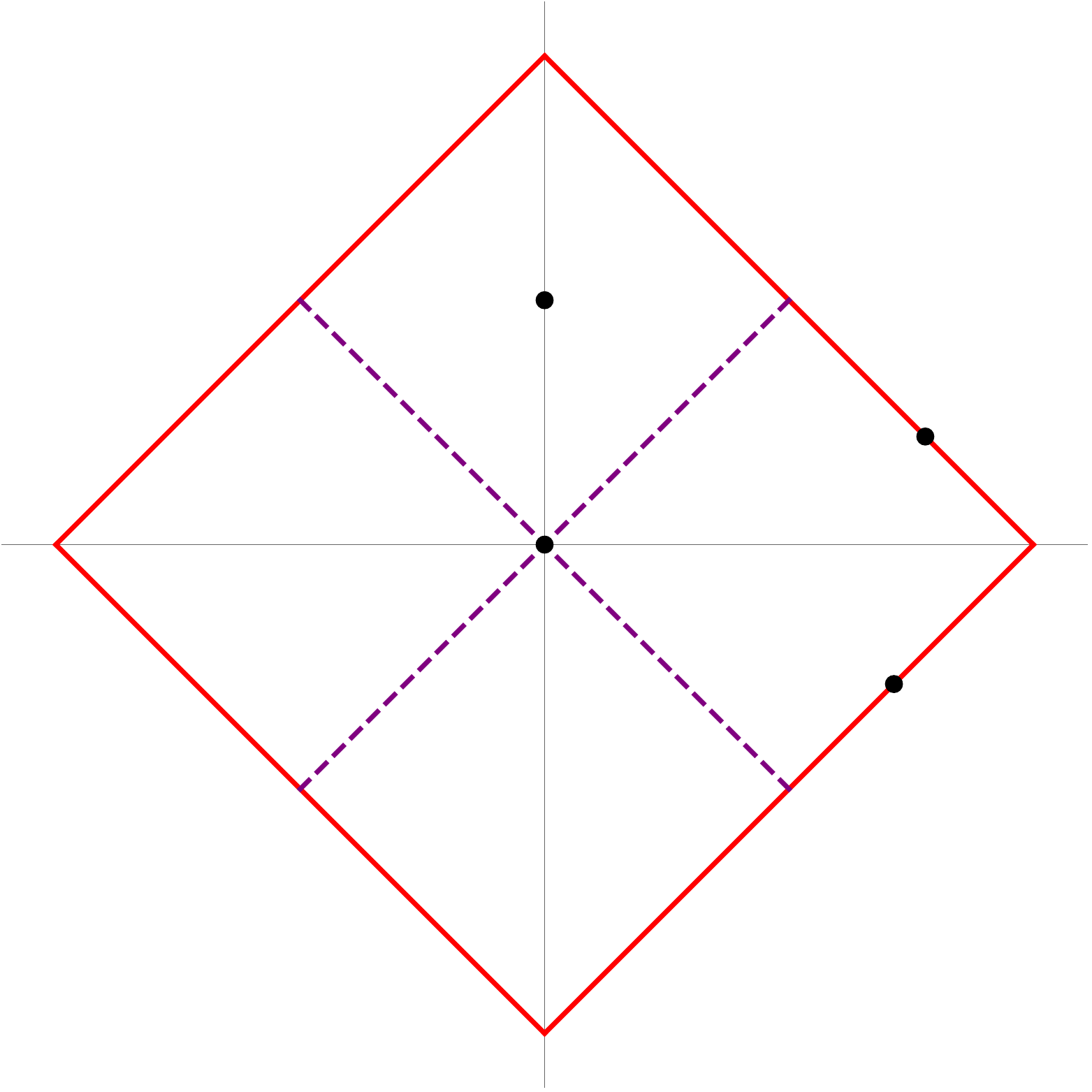}};
    \node at (-.61,0.21){$x_1$};
    \node at (2.8,-1.3){$x_2$};
    \node at (-.4,1.8){$x_3$};
    \node at (3,1){$x_4$};
\end{tikzpicture}
    \end{minipage}%
    \begin{minipage}{0.5\textwidth}
        \begin{tikzpicture}[thick,scale=0.7, every node/.style={scale=0.8}]

\node[inner sep=0pt] (russell) at (0,0)
    {\includegraphics[width=.8\textwidth]{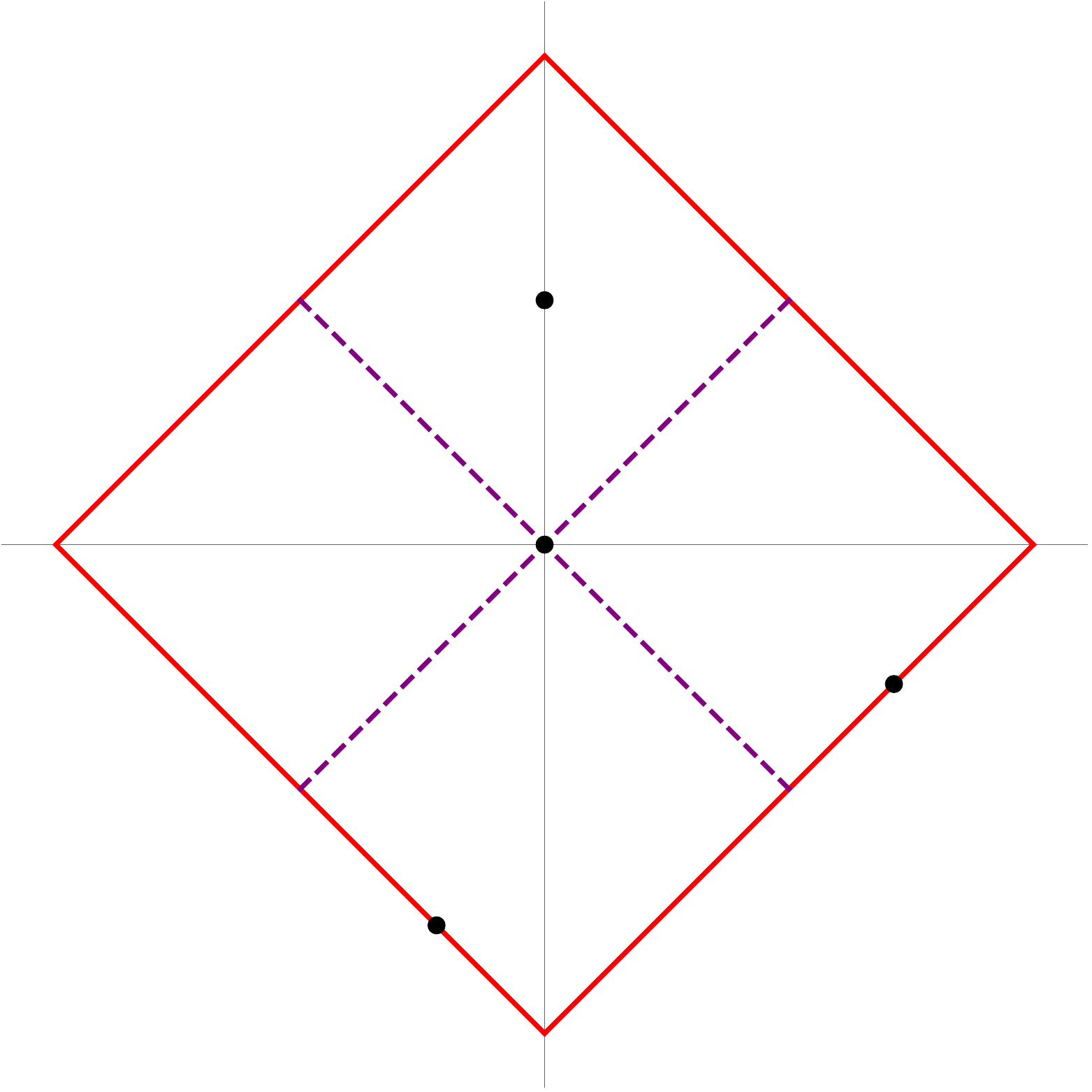}};
    \node at (-.61,0.21){$x_1$};
    \node at (2.8,-1.1){$x_2$};
    \node at (-.4,1.8){$x_3$};
    \node at (-1.2,-2.6){$x_4$};
\end{tikzpicture}
    \end{minipage}
    \caption{Left: Configuration  $\boldsymbol{X}^{+-}$ --- $x_4$ and $x_2$ on $\mathscr{I}^+(\theta=0)$ and $\mathscr{I}^-(\theta=0)$. Right: Configuration  $\boldsymbol{X}^{--}$ --- $x_4$ and $x_2$ on $\mathscr{I}^-(\theta=\pi)$ and $\mathscr{I}^-(\theta=0)$. In terms of unitary matrices $\boldsymbol{X}^{--}$ and $\boldsymbol{X}^{+-}$ coincide due to the identification $\mathscr{I}^+\sim \mathcal{A}\mathscr{I}^-$ (see section \ref{confcompact}).}
    \label{figg20}
\end{figure}
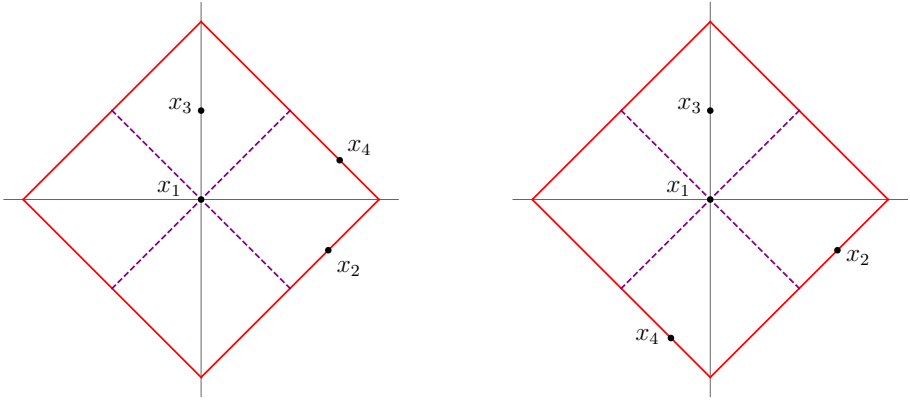

In this section we demonstrate that the integral (\ref{5b}) can be computed for most kinematic regions with real $z,\bar{z}$ by considering special `double infinity' configurations. We consider eight configurations $\boldsymbol{X}^{ab}, \boldsymbol{Y}^{ab}$ of four points in Minkowski space, where $a,b\in\{+,-\}$, defined by the Hermitian matrices (related to usual Minkowski vectors by (\ref{30a}))
\begin{align}\label{1c}
    X_1^{ab}=\begin{pmatrix}0&0\\0&0 \end{pmatrix},\spa \spa  X_2^{ab}=\begin{pmatrix}\xi_+&0\\0&\eta_b \end{pmatrix},\spa\spa    X_3^{ab}=\begin{pmatrix}1&0\\0&1 \end{pmatrix},   \spa\spa X_4^{ab}=\begin{pmatrix}\eta_a&0\\0&\xi_- \end{pmatrix},
\end{align}
\begin{align}\label{2c}
    Y_1^{ab}=\begin{pmatrix}0&0\\0&0 \end{pmatrix},\spa \spa  Y_2^{ab}=\begin{pmatrix}\xi_+&0\\0&-\eta_b \end{pmatrix},\spa\spa    Y_3^{ab}=\begin{pmatrix}1&0\\0&-1 \end{pmatrix},   \spa\spa Y_4^{ab}=\begin{pmatrix}\eta_a&0\\0&-\xi_- \end{pmatrix},
\end{align}
where $\xi_+,\xi_-\in \mathbb{R}\setminus\{0,1\}$ and we take the limit $\eta_\pm\rightarrow\pm\infty$. For each $a,b$ we see that  $X_1^{ab}$ and $Y_1^{ab}$ correspond to the origin in Minkowski space, while $X_3^{ab}$ and $Y_3^{ab}$ correspond to the unit vectors $e_0=(1,0,0,0)$ and $e_3=(0,0,0,1)$ respectively. The remaining points live on $\mathscr{I}^+$ or $\mathscr{I}^-$ depending on the choice of $a$ or $b$, and are parametrised by a degree of freedom $\xi_\pm$. The configurations $\boldsymbol{X}^{+-}$ and $\boldsymbol{X}^{--}$ are visualised on an extended Penrose diagram in figure \ref{figg20}. 
\noindent In the limit $\eta_\pm\rightarrow \pm\infty$ $\boldsymbol{X}^{ab}$ correspond to the same configuration in terms of unitary matrices $\boldsymbol{U}_{\boldsymbol{X}}$, and similarly $\boldsymbol{Y}^{ab}$ correspond to the same configuration $\boldsymbol{U}_{\boldsymbol{Y}}$
\begin{align}\label{1t}
    \boldsymbol{U}_{\boldsymbol{X}}=\left(\begin{pmatrix}1&0\\0&1\end{pmatrix},\begin{pmatrix}\frac{i-\xi_+}{i+\xi_+}&0\\0&-1\end{pmatrix},\begin{pmatrix}i&0\\0&i\end{pmatrix},\begin{pmatrix}-1&0\\0&\frac{i-\xi_-}{i+\xi_-}\end{pmatrix}\right),
\end{align}
\begin{align}
    \boldsymbol{U}_{\boldsymbol{Y}}=\left(\begin{pmatrix}1&0\\0&1\end{pmatrix},\begin{pmatrix}\frac{i-\xi_+}{i+\xi_+}&0\\0&-1\end{pmatrix},\begin{pmatrix}i&0\\0&-i\end{pmatrix},\begin{pmatrix}-1&0\\0&\frac{i+\xi_-}{i-\xi_-}\end{pmatrix}\right).
\end{align}
Note that even though each $\boldsymbol{X}^{ab}$ corresponds to the same configuration in terms of unitary matrices $\boldsymbol{U}_{\boldsymbol{X}}$, the value of the box integral $I(\boldsymbol{X}^{ab})$ depends on $a,b$, i.e.\ the direction $X_2$ and $X_4$ are sent to $\mathscr{I}$ (and similarly for $\boldsymbol{Y}^{ab}$). This is because the box integral is singular for configurations with points on $\mathscr{I}$, and we should think of $X_2,\sp Y_2$ and $X_4,\sp Y_4$ as being slightly off $\mathscr{I}$ so that $\mathscr{I}^+$ and $\mathscr{I}^-$ are distinguished. We take the kinematics to be large, but finite. We compute the kinematics $k(\boldsymbol{X}^{ab})$ and $k(\boldsymbol{Y}^{ab})$ in the limit $\eta_\pm\rightarrow \pm\infty$ to be
\begin{align}
   k(\boldsymbol{X}^{ab})=\left(\sp\xi_+\eta_b,\sp\sp \eta_a(\xi_--1),\sp\sp \eta_b(\xi_+-1),\sp\sp\xi_-\eta_a,\sp\sp 1,\sp\sp-\eta_a\eta_b\sp\right),
\end{align}
\begin{align}\label{3c}
k(\boldsymbol{Y}^{ab})=-k(\boldsymbol{X}^{ab}).
\end{align}
We also compute the conformal cross-ratios in the limit
\begin{align}
    u(\boldsymbol{X}^{ab})= u(\boldsymbol{Y}^{ab})=\xi_+(1-\xi_-),\spac    v(\boldsymbol{X}^{ab})=  v(\boldsymbol{Y}^{ab})=\xi_-(1-\xi_+),
\end{align}
so that the cross-ratios $u$ and $v$ are the same for each of the eight configurations $\boldsymbol{X}^{ab}$ and $\boldsymbol{Y}^{ab}$. Written in terms of $z$ and $\bar{z}$ the cross-ratios for each configuration are
\begin{align}
    z=\text{min}(\xi_+,1-\xi_-),\spac \bar{z}=\text{max}(\xi_+,1-\xi_-),
\end{align}
since by convention we take $\bar{z}\geq z$. Note that since $\xi_+,\xi_-\in \mathbb{R}\setminus\{0,1\}$, the configurations (\ref{1c}) and (\ref{2c}) can be tuned to give any real $z$ and $\bar{z}$ with $\bar{z}\geq z$. Depending on $\xi_+,\sp \xi_-$ and the choice of configuration $\boldsymbol{X}^{ab}$ or $\boldsymbol{Y}^{ab}$ the kinematics of the configuration can vary widely. In fact most configurations $\boldsymbol{w}$ with real $z$ and $\bar{z}$ and a given $\text{sgn}(k(\boldsymbol{w}))$ can be conformally mapped to one of the double infinity configurations $\boldsymbol{w}_\infty$, i.e.\ (\ref{1c}) or (\ref{2c}) (possibly permuted) with the same $z$ and $\bar{z}$ and the \textit{same signs of kinematics} so that $I(\boldsymbol{w})=I(\boldsymbol{w}_\infty)$ by pseudo-conformal invariance. Specialising to the restricted kinematics considered in tables \ref{table8}$-$\ref{table11}, the only kinematic regions which cannot be realised by a double infinity configuration are the missing kinematic region $k^*$ described in section \ref{branches}, and the corresponding region obtained by reversing all kinematics $Pk^*$ (which can be realised by finite configurations). For example, we show how to realise all kinematic regions in $V_2$ in table \ref{table77}.
\begin{table}[h!]
\begin{center}
 \begin{tabular}{||c|c|c|c||} 
 \hline
 Kinematic region $k$ & $\boldsymbol{w}_\infty$ & $\xi_+,\xi_-$ range \\ [0.5ex] 
 \hline
$(-++++\hspace{0.1cm}+)$& $\boldsymbol{X}^{+-}$ & $\xi_+\in(0,1),\sp\xi_-\in(1,\infty)$\\
\hline
$(+++--\hspace{0.1cm}+)$&$\boldsymbol{Y}^{++}$ &$\xi_+\in(-\infty,0),\sp \xi_-\in(0,1)$\\
\hline
$(++-++\hspace{0.1cm}-)$&$\boldsymbol{X}^{++}$ &$\xi_+\in(0,1),\sp\xi_-\in(1,\infty)$\\
\hline
$(---+-\hspace{0.1cm}+)$&$\boldsymbol{Y}^{--}$ &$\xi_+\in(-\infty,0),\sp \xi_-\in(0,1)$\\
\hline
$(--+-+\hspace{0.1cm}-)$&$\boldsymbol{X}^{--}$ &$\xi_+\in(0,1),\sp\xi_-\in(1,\infty)$\\
\hline
$(-+---\hspace{0.1cm}-)$&$\boldsymbol{Y}^{+-}$ &$\xi_+\in(-\infty,0),\sp \xi_-\in(0,1)$\\
\hline
$(+---+\hspace{0.1cm}+)$&$\boldsymbol{X}^{-+}$ &$\xi_+\in(0,1),\sp\xi_-\in(1,\infty)$\\
\hline
$(+-++-\hspace{0.1cm}-)$&$\boldsymbol{Y}^{-+}$ &$\xi_+\in(-\infty,0),\sp \xi_-\in(0,1)$\\
\hline

\end{tabular}
\caption{How to realise all kinematic regions in $V_2$ ($z\in(-\infty,0),\bar{z}\in(0,1)$) with double infinity configurations. No need for permutations in this case.}
\label{table77}
\end{center}
\end{table}What makes these $\boldsymbol{w}_\infty$ configurations particularly interesting is that the box integral (\ref{5b}) with these configurations can be calculated directly in Minkowski space, i.e.\ without reference to Feynman parameters or a Mellin representation. For each configuration the integral depends only on the choice of $\xi_+$ and $\xi_-$ and so we define
\begin{align}
    I^{ab}_{\boldsymbol{X}}(\xi_+,\xi_-)\equiv I(\boldsymbol{X}^{ab}),\spac  I^{ab}_{\boldsymbol{Y}}(\xi_+,\xi_-)\equiv I(\boldsymbol{Y}^{ab}),
\end{align}
where how we make sense of the limit $\eta_{\pm}\rightarrow\pm\infty$ in the calculation will be described shortly. First we note that we can use symmetries to reduce the number of integrals to compute from eight to two. We can immediately combine (\ref{8b}) and (\ref{3c}) to conclude
\begin{align}\label{4c}
    I^{ab}_{\boldsymbol{Y}}(\xi_+,\xi_-)=I^{ab}_{\boldsymbol{X}}(\xi_+,\xi_-)^*.
\end{align}
We can also note relations between the configurations (\ref{1c}) using permutations, discrete transformations, and translations. We have
\begin{align}\label{5c}
    \mathcal{P}\circ(24)\boldsymbol{X}^{+-}=\boldsymbol{X}^{-+} (\xi_+\leftrightarrow \xi_-),
\end{align}
\begin{align}\label{6c}
    T_{\mathbb{I}_2}\circ\mathcal{P}\mathcal{T}\circ(13)\boldsymbol{X}^{--}=\boldsymbol{X}^{++} (\xi_\pm \rightarrow 1-\xi_\pm).
\end{align}
Since the box integral is fully invariant under each of the transformations used in (\ref{5c}) and (\ref{6c}) we see that
\begin{align}\label{7c}
    I^{-+}_{\boldsymbol{X}}(\xi_+,\xi_-)=I^{+-}_{\boldsymbol{X}}(\xi_-,\xi_+),\spac  I^{++}_{\boldsymbol{X}}(\xi_+,\xi_-)=I^{--}_{\boldsymbol{X}}(1-\xi_+,1-\xi_-). 
\end{align}
In section \ref{doubleinfcalc} we describe our results for $I^{+-}_{\boldsymbol{X}}(\xi_+,\xi_-)$ and $I^{--}_{\boldsymbol{X}}(\xi_+,\xi_-)$, leaving the details of the calculation to appendix \ref{doubleinfcalcdetails}. We can then use (\ref{4c}) and (\ref{7c}) to recover the value of the integral for each of the configurations (\ref{1c}) and (\ref{2c}).

\subsection{Double Infinity Box Integral}\label{doubleinfcalc}
We write the integration variable $x_5$ in spherical coordinates
\begin{align}
    x_5=(t,r\sin\theta\cos\phi,r\sin\theta\sin\phi,r\cos\theta),\spac
    X_5=\begin{pmatrix}t+r\cos\theta&re^{-i\phi}\sin\theta\\re^{i\phi}\sin\theta&t-r\cos\theta\end{pmatrix},
\end{align}
with $r>0, \phi\in[0,2\pi), \theta\in[0,\pi]$. We then consider the two cases described above. We consider the configuration $\boldsymbol{X}^{+-}$ defined in (\ref{1c}). The Lorentz squares $x_{i5}^2$ for $i=1,2,3,4$ are calculated in the appropriate limit, for example
\begin{align}
    x_{25}^2=|X_2-X_5|=\det\begin{pmatrix}\xi_+-t-r\cos\theta&-re^{-i\phi}\sin\theta\\-re^{i\phi}\sin\theta&\eta_--t+r\cos\theta\end{pmatrix}
\end{align}
\begin{align*}
    =\eta_-(\xi_+-t-r\cos\theta)+O(\eta_-^0).
\end{align*}
Letting $x=\cos\theta$, for this configuration the box integral becomes $I_{\boldsymbol{X}}^{+-}(\xi_+,\xi_-)\simeq$
\begin{align}\label{12i}
    \frac{2i}{\pi}\int_{x,r,t} \frac{r^2 \eta_+\eta_-}{(t^2-r^2+i\epsilon)((t-1)^2-r^2+i\epsilon)\eta_-(\xi_+-t-rx+\frac{i\epsilon}{\eta_-})\eta_+(\xi_--t+rx+\frac{i\epsilon}{\eta_+})},
\end{align}
where $\int_{x,r,t}\equiv\int_{-1}^1 dx\int_{0}^{\infty}dr\int_{-\infty}^\infty dt $ and the trivial $\phi$ integration has been performed. Note that in the limit $\eta_{\pm}\rightarrow \pm\infty$ the integral is well-defined, as factors of $\eta_\pm$ cancel in numerator and denominator. We define an $\epsilon'$ by $\frac{\epsilon}{\eta_{\pm}}\sim\pm \epsilon'$, and throughout the calculation we treat $\epsilon'\ll\epsilon$ in the limit $\eta_{\pm}\rightarrow \pm\infty$. This is essential to properly regulate the poles of the integrand. We describe the details of the remaining calculation in appendix \ref{doubleinfcalcdetails}. The answer can be expressed in terms of a set of four `basis' integrals
\begin{align}
    h_1(a,b)\equiv (a-b)\int_0^1dx\frac{1}{(x-a)(x-b)},
\end{align}
\begin{align}
    h_2(a,b)\equiv (a-b)\int_0^1dx\frac{\log(x)}{(x-a)(x-b)},
\end{align}
\begin{align}
        h_3(a,b)\equiv (a-b)\int_0^1dx\frac{\log(1-x)}{(x-a)(x-b)},
\end{align}
\begin{align}
        h_4(a,b)\equiv (a-b)\int_0^1dx\frac{\log(1+x)}{(x-a)(x-b)},
        \end{align}
which are all expressible as simple combinations of logs and dilogs of $a$ and $b$, found easily, e.g.\ using 
\texttt{Mathematica}. These are all well-defined in our case as each $a,b$ will have a small imaginary part. The final result for the integral is
\begin{align}\label{7i}
    I_{\boldsymbol{X}}^{+-}(\xi_+,\xi_-)=\frac{1}{\Box\xi-1}\Bigg(\log(1/2)(h_1(r_{1+},r_{2+})-h_1(-r_{1-},-r_{2-}))
\end{align}
\begin{align*}
    +\log(-1/2-i\epsilon)(-h_1(-r_{1+},-r_{2+})+h_1(r_{1-},r_{2-}))-\log(-\Delta\xi/2+i\epsilon)h_1(s_{1+}^2,s_{2+}^2)
\end{align*}
\begin{align*}
    +\log(\Delta\xi/2-i\epsilon)h_1(s_{1-}^2,s_{2-}^2)+\frac{1}{2}(h_2(s_{1+}^2,s_{2+}^2)-h_2(s_{1-}^2,s_{2-}^2))+\Big(\log(\xi_+-i\epsilon)\times
\end{align*}
\begin{align*}
(h_1(-r_{1-},-s_{1-}) -h_1(r_{1+},s_{1+}))+\log(-\xi_--i\epsilon)(h_1(r_{2-},s_{1-})-h_1(-r_{2+},-s_{1+}))
\end{align*}
\begin{align*}
   +h_3(r_{1+},s_{1+})+h_3(-r_{2+},-s_{1+})-h_4(-r_{1-},-s_{1-})-h_4(r_{2-},s_{1-})-(\xi_{\pm}\rightarrow 1-\xi_{\mp})\Big)\Bigg),
\end{align*}
where $\Box\xi=\xi_++\xi_-, \Delta\xi=\xi_+-\xi_-$, and $r_{i\pm}$ and $s_{i\pm}$ for $i=1,2$ are functions of $\xi_+,\xi_-$ defined in (\ref{u1}) and (\ref{u2}). (\ref{7i}) is invariant under $\xi_\pm\rightarrow 1-\xi_\mp$. Note that the overall prefactor agrees with our expectation
\begin{align}
    \frac{1}{\Box\xi-1}=\pm\frac{1}{z-\bar{z}}.
\end{align}
For the configuration $\boldsymbol{X}^{--}$ the integral becomes $ I_{\boldsymbol{X}}^{--}(\xi_+,\xi_-)=$
\begin{align}
 \frac{2i}{\pi}\int_{x,r,t} \frac{r^2}{(t^2-r^2+i\epsilon)((t-1)^2-r^2+i\epsilon)(\xi_+-t-rx-i\epsilon')(\xi_--t+rx-i\epsilon')}.
\end{align}
The final result for the integral is
\begin{align}\label{8i}
    I_{\boldsymbol{X}}^{--}(\xi_+,\xi_-)=\frac{1}{\Box\xi-1}\Bigg(\log(1/2)(h_1(r_{1+},r_{2+})-h_1(-r_{1-},-r_{2-}))+\log(-1/2-i\epsilon) \times
\end{align}
\begin{align*}
   (h_1(r_{1-},r_{2-})-h_1(-r_{1+},-r_{2+}))+\Big(-\log(\xi_+-i\epsilon)h_1(r_{1+},s_{1+}) +\log(\xi_--i\epsilon)h_1(r_{2+},s_{1+})
    \end{align*}
\begin{align*}
   +\log(-1+\xi_+-i\epsilon)h_1(-r_{1+},s_{2+})+\log(-1+\xi_--i\epsilon)h_1(-r_{2+},s_{2+})
\end{align*}
\begin{align*}
   +h_3(r_{1+},s_{1+})-h_3(-r_{1+},s_{2+})+h_4(-r_{2+},s_{2+})-h_4(r_{2+},s_{1+})+(\xi_{+}\leftrightarrow \xi_{-})\Big)\Bigg),
\end{align*}
and is invariant under $\xi_+\leftrightarrow\xi_-$. Using (\ref{4c}) and (\ref{7c}) an expression for the box integral (\ref{5b}) for each of the eight configurations in (\ref{1c}) and (\ref{2c}) can be recovered from (\ref{7i}) and (\ref{8i}). These results were numerically verified in each accessible kinematic region with tables \ref{table8}$-$\ref{table11} as well as with the package \cite{vanOldenborgh:1990yc}. Note there are numerical issues when $\Delta\xi=0,\Box\xi=0, 1, 2$, $\xi_+=1/2,$ or $\xi_-=1/2$. These are easily remedied however by adding a small correction to $\xi_+$ or $\xi_-$.                                                               

\subsection{Numerical Example}\label{numeric}
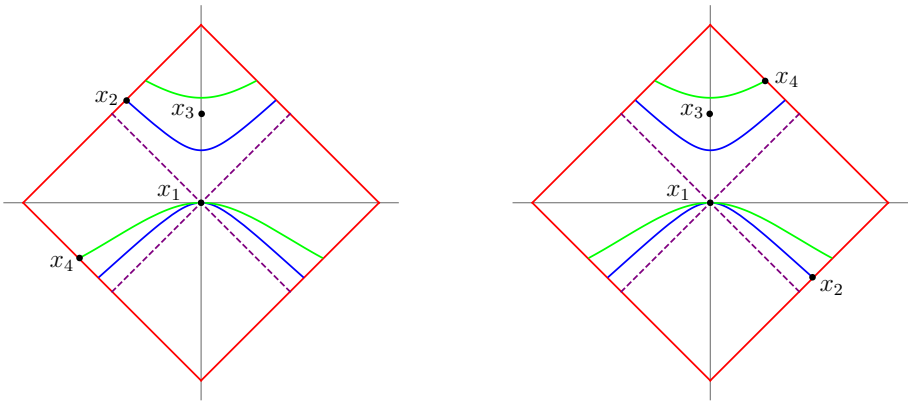
\begin{figure}[!htb]
    \centering
    \begin{minipage}{.5\textwidth}
        \centering
       \begin{tikzpicture}[thick,scale=0.7, every node/.style={scale=0.8}]

\node[inner sep=0pt] (russell) at (0,0)
    {\includegraphics[width=.8\textwidth]{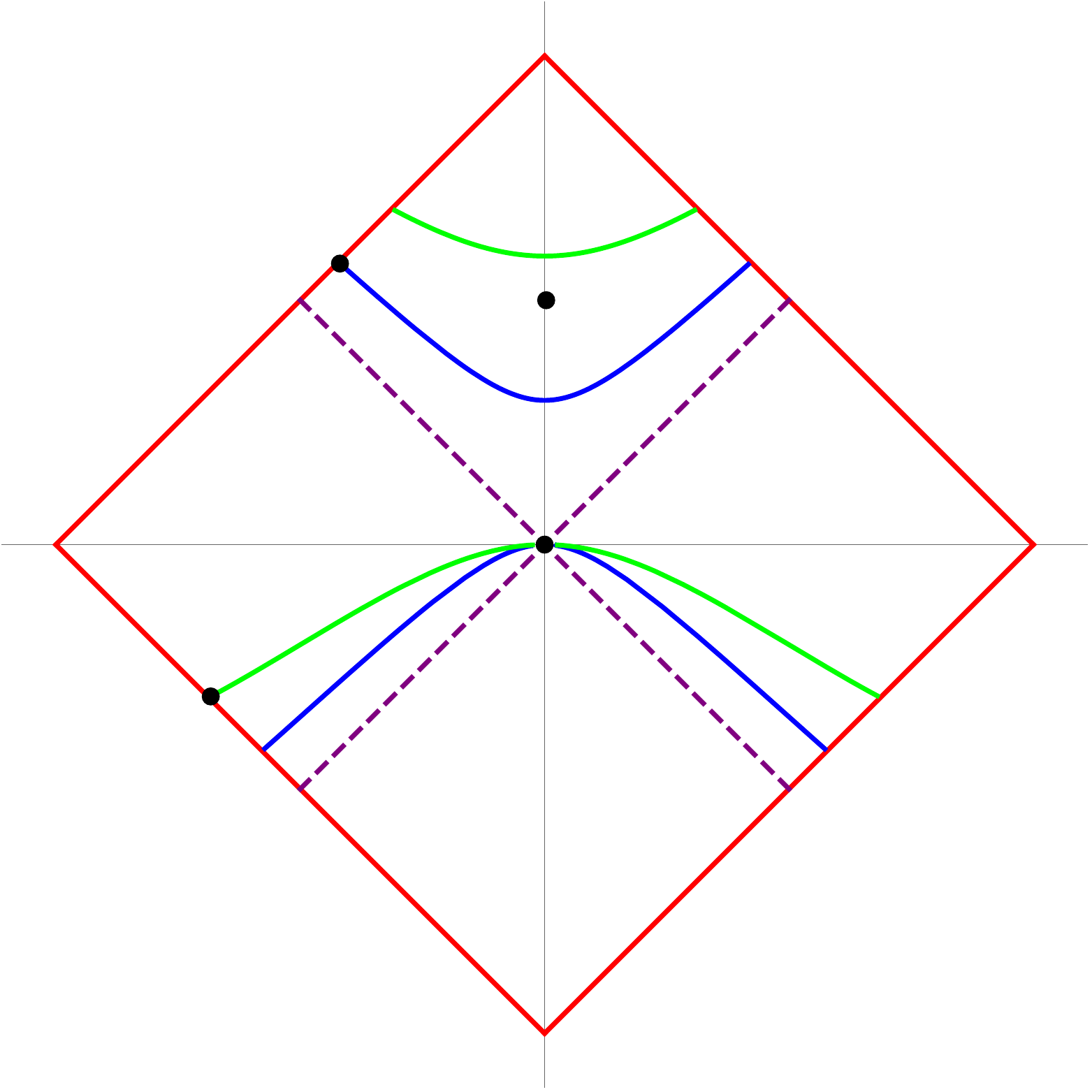}};
    \node at (-.61,0.21){$x_1$};
    \node at (-1.8,2){$x_2$};
    \node at (-.35,1.75){$x_3$};
    \node at (-2.65,-1.15){$x_4$};
\end{tikzpicture}
    \end{minipage}%
    \begin{minipage}{0.5\textwidth}
        \begin{tikzpicture}[thick,scale=0.7, every node/.style={scale=0.8}]

\node[inner sep=0pt] (russell) at (0,0)
    {\includegraphics[width=.8\textwidth]{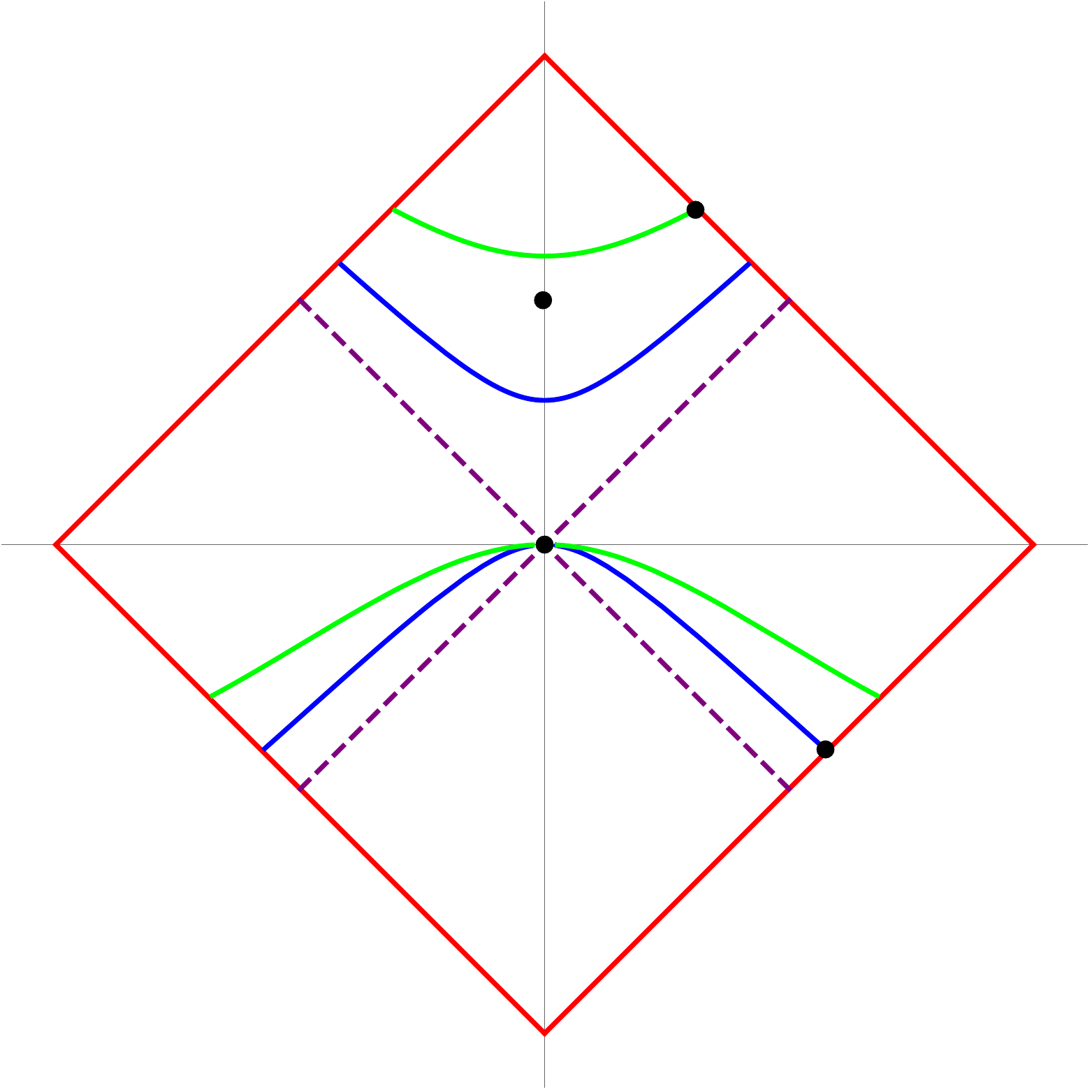}};
    \node at (-.61,0.21){$x_1$};
    \node at (2.3,-1.6){$x_2$};
    \node at (-.35,1.75){$x_3$};
    \node at (1.45,2.35){$x_4$};
\end{tikzpicture}
    \end{minipage}
    \caption{Configurations $\boldsymbol{w}$ (left) and $\boldsymbol{y}$ (right). The blue and green curves show the trajectories of $x_2$ and $x_4$ respectively under the SCT $C_{(0,0,0,b)}$ for $b\in\mathbb{R}$. $x_4$ crosses infinity again at $b=3/2+\epsilon$.}
    \label{figg28}
\end{figure}\noindent As a concrete example, consider the configuration $\boldsymbol{w}\equiv \boldsymbol{X}^{-+}(\xi_+=1/4,\xi_-=2/3)$. For this configuration we have $z=1/4,\bar{z}=1/3$, $\text{sgn}(k(\boldsymbol{w}))=(++--+\hspace{0.11cm}+)$, and $\boldsymbol{w}\in \bar{V}_1$. The value of the box integral is $I^{-+}_{\boldsymbol{X}}(1/4,2/3)\simeq 5.88-21.69i$, which can be calculated using (\ref{7i}) and (\ref{7c}), table \ref{table9}, or for example the package \cite{vanOldenborgh:1990yc}. Under the infinitesimal SCT $C_{(0,0,0,\epsilon)}$ $\boldsymbol{w}$ is mapped to $\boldsymbol{y}\equiv\boldsymbol{X}^{+-}(\xi_+=1/4,\xi_-=2/3)$ with $\text{sgn}(k(\boldsymbol{y}))=(--+++\hspace{0.05cm}+)$ and $I^{+-}_{\boldsymbol{X}}(1/4,2/3)\simeq 5.88-8.88i$. Under the SCT $C_{(0,0,0,3/2+\epsilon)}$  $\boldsymbol{w}$ is mapped to a configuration $\boldsymbol{z}$ with $\text{sgn}(k(\boldsymbol{z}))=(-+-+-\hspace{0.12cm}+)$ and $I(\boldsymbol{z})\simeq 5.88$. This example shows three of the four branches of the box integral accessible in $\bar{V}_{1,1/4,1/3}$ (table \ref{table9}). In figure \ref{figg28} we show the configurations $\boldsymbol{w}$ and $\boldsymbol{y}$ and the orbits of $x_2$ and $x_4$ under the SCT parametrised by $(0,0,0,b)$.

\section{Conclusions}
Due to the causality-violating nature of the conformal group, the one-loop massless box integral is not fully conformally invariant in Minkowski space. We classified conformally equivalent configurations of four points in compactified Minkowski space, and found a much richer structure than the corresponding Euclidean classification. In Minkowski space the conformal invariants $z$ and $\bar{z}$ are not enough to decide if two configurations are conformally equivalent, and further information about the kinematic invariants $x_{ij}^2$ is needed. It is worth noting that a result similar in spirit appeared recently in \cite{qiao2020classification}, although the classifications are slightly different and serve different purposes. We rewrote the result \cite{Duplan_i__2002} for the box integral explicitly in terms of the conformal invariants $z$ and $\bar{z}$ in each kinematic region. Combining this and our classification of conformally equivalent configurations we conclude that using SCTs up to four branches of the box integral can be reached starting from any configuration, and up to six values of the integral for a given $z,\bar{z}$. There are two qualitatively different behaviours for the branch structure, one which appears in tables \ref{table8},\ref{table9} and one in tables \ref{table10},\ref{table11}. One interesting fact is that the set $V_1$ of conformally equivalent configurations contains only \textit{three} branches of the box integral. This is due to the `missing' kinematic region, the non-existence of which in general we were so far only able to establish numerically. We studied a range of double infinity configurations in Minkowski space, and found that the vast majority of kinematic regions for configurations with $z,\bar{z}\in\mathbb{R}$ can be realised. We were able to calculate the box integral for these configurations directly in Minkowski space. By pseudo-conformal invariance we can use this calculation to find the value of the box integral for all finite configurations with real $z,\bar{z}$ except for a single special case (the kinematics reversed version of the missing region, which is numerically found to be rare). One disadvantage of the results (\ref{7i}) and (\ref{8i}) is their relative non-transparency. While they are numerically easy to work with, it is technically difficult to analytically reduce them to the form (\ref{4i}) for the different kinematic regions, although we found total numerical agreement. In any case it is interesting that it is even possible to calculate the box integral directly in Minkowski space for such a large class of kinematics.
\newline
\newline
There are a number of interesting directions for further research. Since the box integral is the simplest of a large family of (locally) dual conformal integrals, one could study the extent to which dual conformal invariance is broken in other examples. We expect the situation to be similar for other four-point conformal integrals, although the number of branches reachable via SCTs may change. At higher points it would be interesting to see how the classification of conformally equivalent configurations changes, and what kind of missing kinematic regions appear. Some questions have been answered in the case of null hexagon configurations in \cite{dorn2012conformal}. At four points we found that the kinematic regions which are not realised by double infinity configurations can be realised by special triple and quadruple infinity configurations \cite{workinprogress}. These configurations are interesting because together with the double infinity configurations they can generate configurations with any $z,\bar{z}$ and any allowed $\text{sgn}(k(\boldsymbol{w}))$. However, it is much trickier to calculate the box integral in these configurations directly in Minkowski space because of nontrivial $\phi,\sin\theta$ dependence and square roots in denominators. Therefore we are unsure if such configurations can be useful for calculations. It would be interesting to study `higher infinity' configurations at higher points and check if all possible kinematic regions can always be realised.

\section*{Acknowledgements}
We thank Lance Dixon, Julian Miczajka, and Lorenzo Quintavalle for helpful discussions. Thanks to Julian and Lorenzo for comments on the manuscript. MS thanks Ewha Womans University for hospitality and his host Changrim Ahn for inspiring discussions. MS thanks the Brain Pool Program of the Korean National Research Foundation (NRF) under grant 2-2019-1283-001-1 for generous support. This project has received funding from the  European Union’s Horizon 2020 research and innovation programme under the Marie Sklodowska-Curie grant agreement  No. 764850 “SAGEX”.

\appendix
\section{\texorpdfstring{$SU(2,2)$}{su22} Conformal Transformations}\label{su22}
We list the subgroups of $SU(2,2)$ corresponding to the familiar conformal transformations on Minkowski space. It is useful to introduce the matrix $W=\begin{pmatrix}-\mathbb{I}_2&i\mathbb{I}_2\\ \mathbb{I}_2&i\mathbb{I}_2\end{pmatrix}$. The $SU(2,2)$ subgroups corresponding to translations and SCTs are
\begin{align}
    T\equiv\left\{t(B)=W\begin{pmatrix}\mathbb{I}_2& B\\0&\mathbb{I}_2\end{pmatrix}W^{-1}\sp|\sp B\in H_{2\times 2}\right\}\subset SU(2,2),
\end{align}
\begin{align}
    C\equiv\left\{c(B)=W\begin{pmatrix}\mathbb{I}_2& 0\\\mathcal{T}B&\mathbb{I}_2\end{pmatrix}W^{-1}\sp|\sp B\in H_{2\times 2}\right\}\subset SU(2,2),
\end{align}
where $H_{2\times 2}$ is the set of $2\times 2$ Hermitian matrices. $T$ and $C$ correspond to translations and SCTs respectively and are parametrised by a Minkowski vector $b^\mu=\frac{1}{2}\tr(\si^\mu B)$. With our conventions SCTs act on finite points $x\in \mathbb{R}^{1,3}$ as
\begin{align}\label{u3}
    \mathfrak{C}_{c(B)}:x^\mu\rightarrow\frac{x^\mu-x^2b^\mu}{1-2b\cdot x+b^2x^2}=\frac{x^\mu-x^2b^\mu}{\si_b(x)}.
\end{align}
Lorentz rotations are implemented by the subgroup
\begin{align}
    L\equiv\left\{l(A)=W\begin{pmatrix}A&0\\0&(A^{\dagger})^{-1}\end{pmatrix}W^{-1}\sp|\sp A\in SL(2,\mathbb{C})\right\}\subset SU(2,2).
\end{align}
Dilatations can be implemented by the subgroup
\begin{align}
    D\equiv \left\{d(\lambda)=W\begin{pmatrix}\la^{1/2}\sp\mathbb{I}_2&0\\0&\la^{-1/2}\mathbb{I}_2\end{pmatrix}W^{-1}\sp|\sp \la>0\right\}\subset SU(2,2).
\end{align}
Together $T,C,L,D\subset SU(2,2)$ constitute the usual $15$ parameter group of conformal transformations connected to the identity. The discrete transformations $\mathcal{P},\mathcal{T}$ take a simple form
\begin{align}\label{6j}
    \mathcal{P}:U\rightarrow (\det U)U^{\dagger},\spac  \mathcal{T}:U\rightarrow (\det U^{-1})U,\spac \mathcal{P}\mathcal{T}:U\rightarrow U^{\dagger}.
\end{align}
There are further $SU(2,2)$ matrices which represent elements of $\text{Conf}(\mathbb{R}^{1,3})$ and cannot be expressed as a product of elements in $T,C,L,D$ \cite{1976JMP....17...24P}. One example is the metric itself, which acts as a `time-inversion'. On finite points $x\notin L_0\sp\spa\mathfrak{C}_{K}$ acts as
\begin{align}
\mathfrak{C}_{K}:x\rightarrow \mathcal{T}\mathcal{I}x.
\end{align}
Combining $\mathfrak{C}_{K}$ and (\ref{6j}) one can get an expression for the action of $\mathcal{I}$ on any $U\in U(2)$.

\newpage
\section{Signs of Kinematics}\label{kin}
We list explicitly in table \ref{table7} the elements of $K_i$ and $\bar{K}_i$ for $i=1,2,3,4.$
\begin{table}[h!]
\begin{center}
 \begin{tabular}{||c|c|c|c||} 
 \hline
 $K_1$ & $K_2$ & $K_3$ & $K_4$ \\ [0.5ex] 
 \hline
$(+++++\hspace{0.1cm}+)$&$(-++++\hspace{0.1cm}+)$&$(++-++\hspace{0.1cm}+)$&$(++++-\hspace{0.1cm}+)$\\
$(-++--\hspace{0.1cm}+)$&$(+++--\hspace{0.1cm}+)$&$(-+---\hspace{0.1cm}+)$&$(-++-+\hspace{0.1cm}+)$\\
$(-+-++\hspace{0.1cm}-)$&$(++-++\hspace{0.1cm}-)$&$(-++++\hspace{0.1cm}-)$&$(-+-+-\hspace{0.1cm}-)$\\
$(+--+-\hspace{0.1cm}+)$&$(---+-\hspace{0.1cm}+)$&$(+-++-\hspace{0.1cm}+)$&$(+--++\hspace{0.1cm}+)$\\
$(+-+-+\hspace{0.1cm}-)$&$(--+-+\hspace{0.1cm}-)$&$(+---+\hspace{0.1cm}-)$&$(+-+--\hspace{0.1cm}-)$\\
$(++---\hspace{0.1cm}-)$&$(-+---\hspace{0.1cm}-)$&$(+++--\hspace{0.1cm}-)$&$(++--+\hspace{0.1cm}-)$\\
$(----+\hspace{0.1cm}+)$&$(+---+\hspace{0.1cm}+)$&$(--+-+\hspace{0.1cm}+)$&$(-----\hspace{0.1cm}+)$\\
$(--++-\hspace{0.1cm}-)$&$(+-++-\hspace{0.1cm}-)$&$(---+-\hspace{0.1cm}-)$&$(--+++\hspace{0.1cm}-)$\\
\hline
$\bar{K}_1$ & $\bar{K}_2$ & $\bar{K}_3$ & $\bar{K}_4$ \\ [0.5ex] 
\hline
$(-----\hspace{0.1cm}-)$&$(+----\hspace{0.1cm}-)$&$(--+--\hspace{0.1cm}-)$&$(----+\hspace{0.1cm}-)$\\
$(+--++\hspace{0.1cm}-)$&$(---++\hspace{0.1cm}-)$&$(+-+++\hspace{0.1cm}-)$&$(+--+-\hspace{0.1cm}-)$\\
$(+-+--\hspace{0.1cm}+)$&$(--+--\hspace{0.1cm}+)$&$(+----\hspace{0.1cm}+)$&$(+-+-+\hspace{0.1cm}+)$\\
$(-++-+\hspace{0.1cm}-)$&$(+++-+\hspace{0.1cm}-)$&$(-+--+\hspace{0.1cm}-)$&$(-++--\hspace{0.1cm}-)$\\
$(-+-+-\hspace{0.1cm}+)$&$(++-+-\hspace{0.1cm}+)$&$(-+++-\hspace{0.1cm}+)$&$(-+-++\hspace{0.1cm}+)$\\
$(--+++\hspace{0.1cm}+)$&$(+-+++\hspace{0.1cm}+)$&$(---++\hspace{0.1cm}+)$&$(--++-\hspace{0.1cm}+)$\\
$(++++-\hspace{0.1cm}-)$&$(-+++-\hspace{0.1cm}-)$&$(++-+-\hspace{0.1cm}-)$&$(+++++\hspace{0.1cm}-)$\\
$(++--+\hspace{0.1cm}+)$&$(-+--+\hspace{0.1cm}+)$&$(+++-+\hspace{0.1cm} +)$&$(++---\hspace{0.1cm}+)$\\
\hline
\end{tabular}
\caption{Explicit signs of kinematics in \texorpdfstring{$K_i$}{Ki} and \texorpdfstring{$\bar{K}_i$}{Ki}.}
\label{table7}
\end{center}
\end{table}

\section{Conformal Plane Extras}\label{confplaneex}
\subsection{\texorpdfstring{$V_\mathbb{C}\subset \bar{V}_{1,z,\bar{z}}$}{Vc}}\label{confplanepseudoeuc}
We prove that all configurations $\boldsymbol{U}\in V_{\mathbb{C}}$ are necessarily contained within $\bar{V}_{1,z,\bar{z}}$. Without loss of generality let $\boldsymbol{w}\in V_{\mathbb{C}}$ be finite. By following the arguments leading to ($\ref{8h}$) we can find $A_1\in\text{Conf}(\mathbb{R}^{1,3})$ such that
\begin{align}
    A_1\boldsymbol{w}=(0,y_2,y_3,\iota).
\end{align}
There are eight possibilities for $(\text{sgn}(y_2^2),\text{sgn}(y_3^2),\text{sgn}(y_{23}^2))$, corresponding to the eight possibilities for the signs of kinematics $K_i$ or $\bar{K}_i$. The only sign assignment which can possibly lead to a conformal plane with $z,\bar{z}\in \mathbb{C}\setminus\mathbb{R}$ is $(---)$, which occurs when $\boldsymbol{w}\in \bar{V}_1$. In this case we can rotate/rescale $y_3$ using $A_2\in\text{Conf}(\mathbb{R}^{1,3})$ to be the spacelike unit vector $e_3=(0,0,0,1)$
\begin{align}
    A_2A_1\boldsymbol{w}=(0,z_2,e_3,\iota),
\end{align}
where $z_2=(t,p,q,r)\equiv A_2y_2$. The stabiliser of $0$, $e_3$, $\iota$ is $SO^+(1,2)$ acting on the first three coordinates of $\mathbb{R}^{1,3}$. The final transformation $A_3\in\text{Conf}(\mathbb{R}^{1,3})$ mapping $\boldsymbol{w}$ to a Minkowskian conformal plane depends on the sign of $c\equiv t^2-p^2-q^2$, which is as yet undetermined because $z_2^{\sp2}<0$. If $\text{sgn}(c)>0$ then $\boldsymbol{w}$ can be mapped to $\boldsymbol{w}_{--}(a,\eta)$ (see table \ref{table2}), which has $z,\bar{z}\in\mathbb{R}$. If $\text{sgn}(c)<0$ then $\boldsymbol{w}$ can be mapped to $\boldsymbol{w}_{\mathbb{C}}(r,\phi)$, which has $z,\bar{z}\in \mathbb{C}\setminus\mathbb{R}$. If $(\text{sgn}(y_2^2),\text{sgn}(y_3^2),\text{sgn}(y_{23}^2))$ is not $(---)$ it is always possible to find a permutation $\si\in S_4$ and a conformal transformation $A$ to map $(0,y_2,y_3,\iota)$ to a Minkowskian conformal plane with real $z,\bar{z}$. Therefore $\boldsymbol{w}$ can only have $z,\bar{z}\in\mathbb{C}\setminus\mathbb{R}$ if $\boldsymbol{w}\in\bar{V}_1$, and so $V_{\mathbb{C}}\subset \bar{V}_{1,z,\bar{z}}$.

\subsection{Missing Kinematic Region}\label{missing}
We show our numerical procedure for excluding the possibility of configurations $\boldsymbol{w}\in V_1$ with $\text{sgn}(k(\boldsymbol{w}))=(----+\sp+)$ and $z,\bar{z}\in(0,1)$. We can bring such an $\boldsymbol{w}=(x_1,x_2,x_3,x_4)\in V_1$ to a simpler form using conformal transformations which do not change the signs of the kinematics. We translate $x_1$ to the origin and rotate/rescale $x_2$ to $e_3$. We can then use an $SO^+(1,2)$ transformation to rotate $x_3$ into the $03$ plane. We can finally use an element of $SO(2)$ to eliminate one of the spatial coordinates from $x_4$. The resulting configuration is
\begin{align}
    \boldsymbol{w}_a=(y_1,y_2,y_3,y_4)=\left(0,e_3,\frac{1}{2}(v_3+u_3,0,0,v_3-u_3),\frac{1}{2}(v_4+u_4,0,h,v_4-u_4)\right),
\end{align}
where $u_i,v_i$ are light cone coordinates and $h$ is the residual spatial coordinate of $y_4$. The kinematics of $\boldsymbol{w}_a$ are $k(\boldsymbol{w}_a)=$
\begin{align}
    \left(-1,(u_3-u_4)(v_3-v_4)-\frac{h^2}{4},(1+u_3)(-1+v_3),u_4v_4-\frac{h^2}{4}, u_3v_3,(1+u_4)(-1+v_4)-\frac{h^2}{4}\right).
\end{align}
$\boldsymbol{w}_a$ is subject to the constraint $\text{sgn}(k(\boldsymbol{w}_a))=(----+\hspace{0.1cm}+)$. For $h=0$ $\boldsymbol{w}_a$ is a two-dimensional configuration and it is easy to prove that $z,\bar{z}\in (-\infty,0)$ or $(1,\infty)$. In this case
we have
\begin{align}
 z,\sp\bar{z}=1-\frac{u_4(1+u_3)}{u_3(1+u_4)},\spa 1-\frac{v_4(1-v_3)}{v_3(1-v_4)}.
\end{align}
Imposing the constraints $\text{sgn}(k(\boldsymbol{w}_a))=(----+\hspace{0.1cm}+)$ for $h=0$ one can see that $z,\bar{z}\in (-\infty,0)$ or $(1,\infty)$, in particular $z,\bar{z}\notin(0,1)$. For $h\neq 0$ this fact is checked numerically (figure \ref{figg25}). It was also checked numerically by taking random configurations $\boldsymbol{w}\in V_1$ with $z,\bar{z}\in (0,1)$, making on the order of $10^7$ SCTs to these, and checking $\text{sgn}(k(\boldsymbol{w}))$ after each iteration. Indeed $\text{sgn}(k(\boldsymbol{w}))=(----+\hspace{0.08cm}+)$ was never observed, as expected.
\begin{figure}[!htb]
    \centering
    \begin{minipage}{.5\textwidth}
        \centering
       \begin{tikzpicture}[thick,scale=0.7, every node/.style={scale=0.8}]

\node[inner sep=0pt] (russell) at (0,0)
    {\includegraphics[width=.95\textwidth]{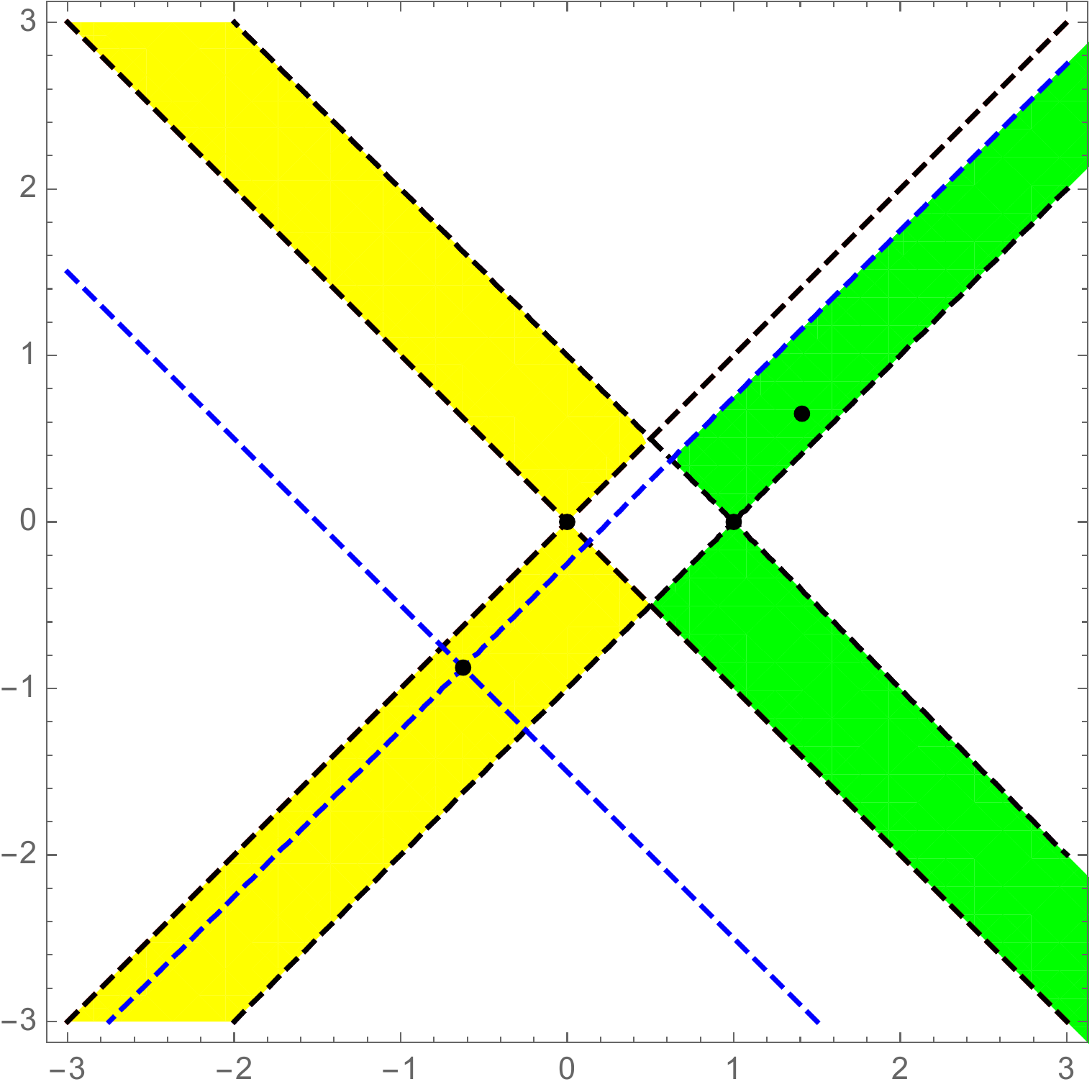}};
 \tiny
    \node at (-.21,0.21){$y_1$};
    \node at (1.15,0.21){$y_2$};
    \node at (-.27,-0.99){$y_3$};
    \node at (1.85,1.1){$y_4$};
    \normalsize
\end{tikzpicture}
    \end{minipage}%
    \begin{minipage}{0.5\textwidth}
        \begin{tikzpicture}[thick,scale=0.7, every node/.style={scale=0.8}]

\node[inner sep=0pt] (russell) at (0,0)
    {\includegraphics[width=.95\textwidth]{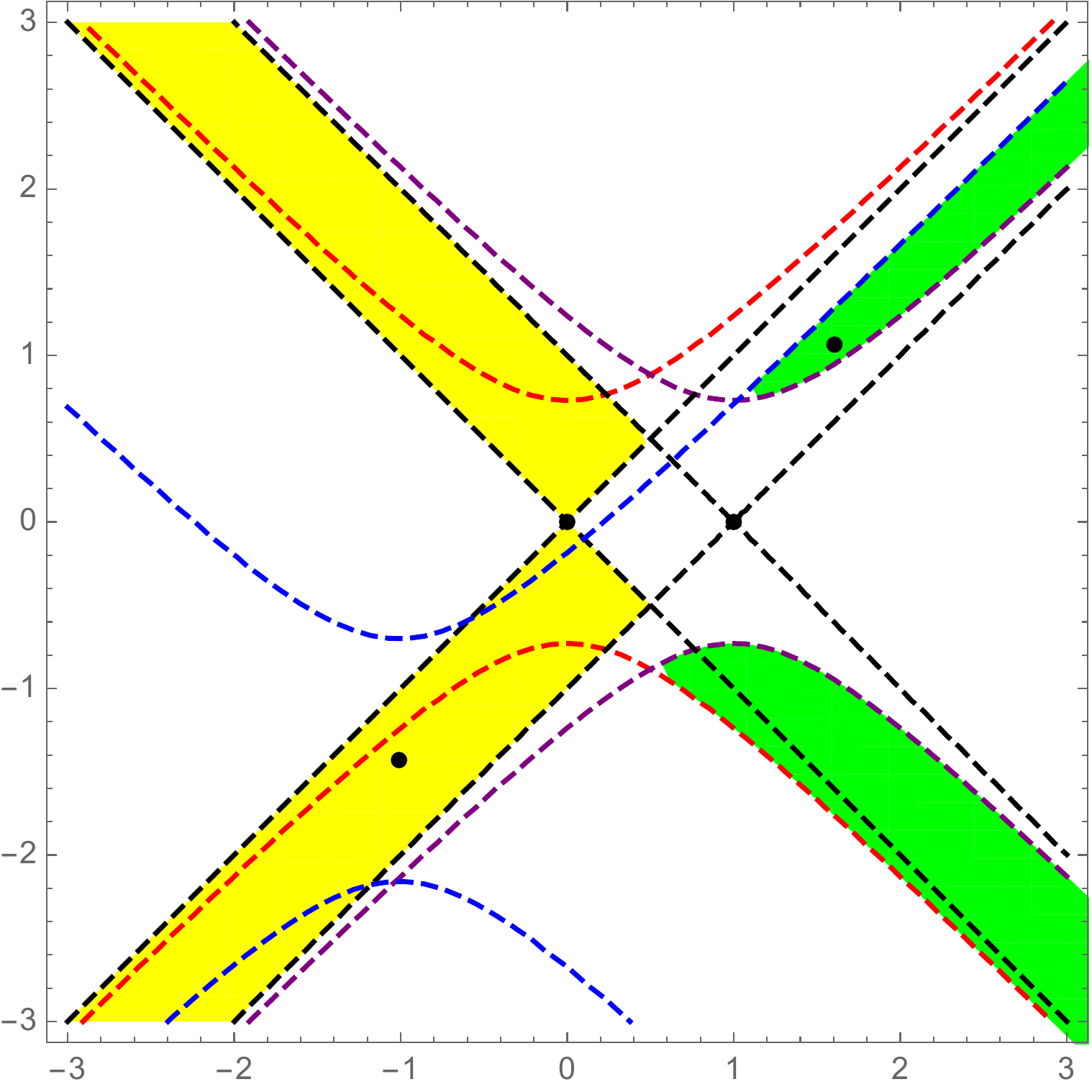}};
   \tiny
    \node at (-.21,0.21){$y_1$};
    \node at (1.15,0.21){$y_2$};
    \node at (-.85,-1.75){$y_3$};
    \node at (2.54,1.85){$y_4$};
    \normalsize
\end{tikzpicture}
    \end{minipage}
    \caption{Configuration $\boldsymbol{w}_a$. $y_3$ can be placed anywhere in yellow region, after which $y_4$ must be placed in green region to enforce $\text{sgn}(k(\boldsymbol{w}_a))=(----+\hspace{0.1cm}+)$. The left picture corresponds to $h=0$ and the right picture to $h\neq 0$. In both cases for this choice of $y_3$ placing $y_4$ in the upper green region gives $z,\bar{z}\in(-\infty,0)$, and placing $y_4$ in lower green region gives $z,\bar{z}\in(1,\infty)$. In the right picture the red, purple, and blue dashed lines correspond to the curves $y_{14}^2=0, y_{24}^2=0,$ and $y_{34}^2=0$ respectively and reduce to light cones at $h=0$.}
    \label{figg25}
\end{figure}
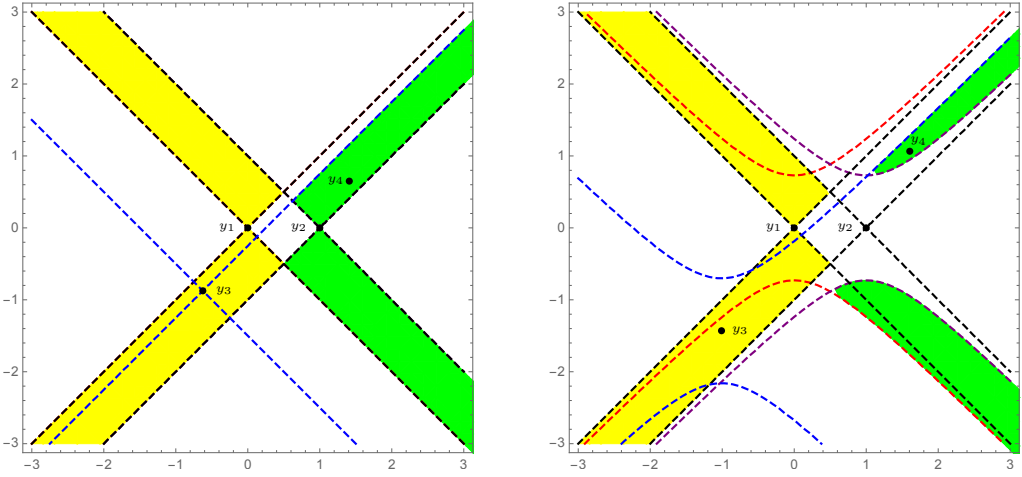
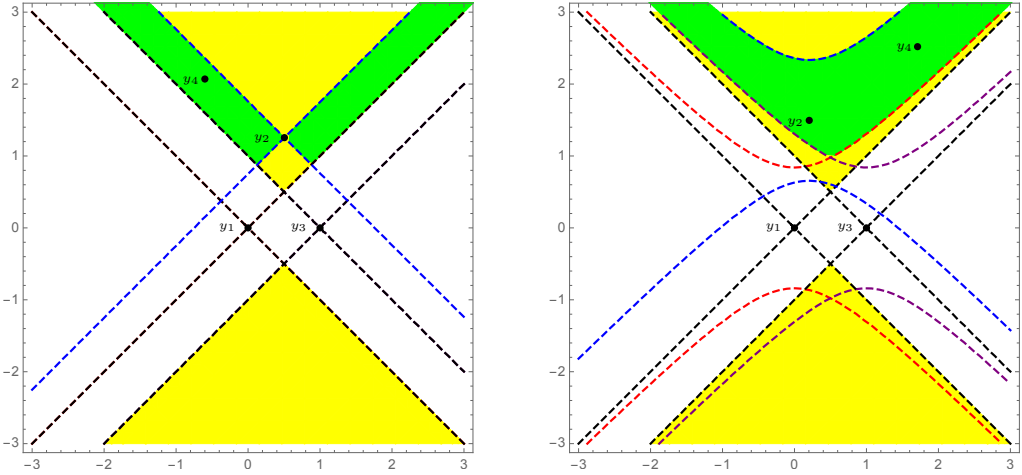
\begin{figure}[!htb]
    \centering
    \begin{minipage}{.5\textwidth}
        \centering
       \begin{tikzpicture}[thick,scale=0.7, every node/.style={scale=0.8}]

\node[inner sep=0pt] (russell) at (0,0)
    {\includegraphics[width=.95\textwidth]{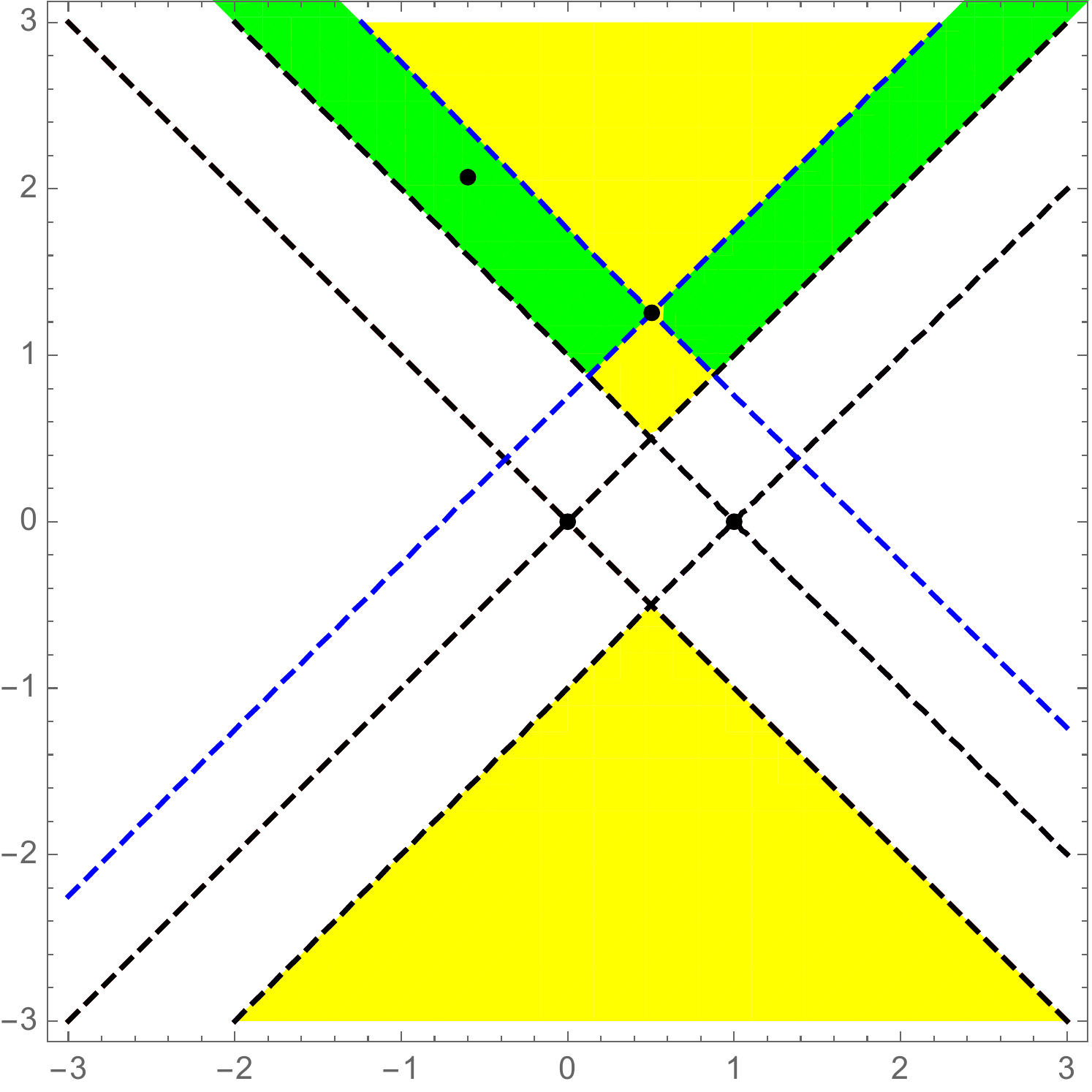}};
 \tiny
    \node at (-.21,0.21){$y_1$};
    \node at (1.15,0.21){$y_3$};
    \node at (0.45,1.9){$y_2$};
    \node at (-0.9,3){$y_4$};
    \normalsize
\end{tikzpicture}
    \end{minipage}%
    \begin{minipage}{0.5\textwidth}
        \begin{tikzpicture}[thick,scale=0.7, every node/.style={scale=0.8}]

\node[inner sep=0pt] (russell) at (0,0)
    {\includegraphics[width=.95\textwidth]{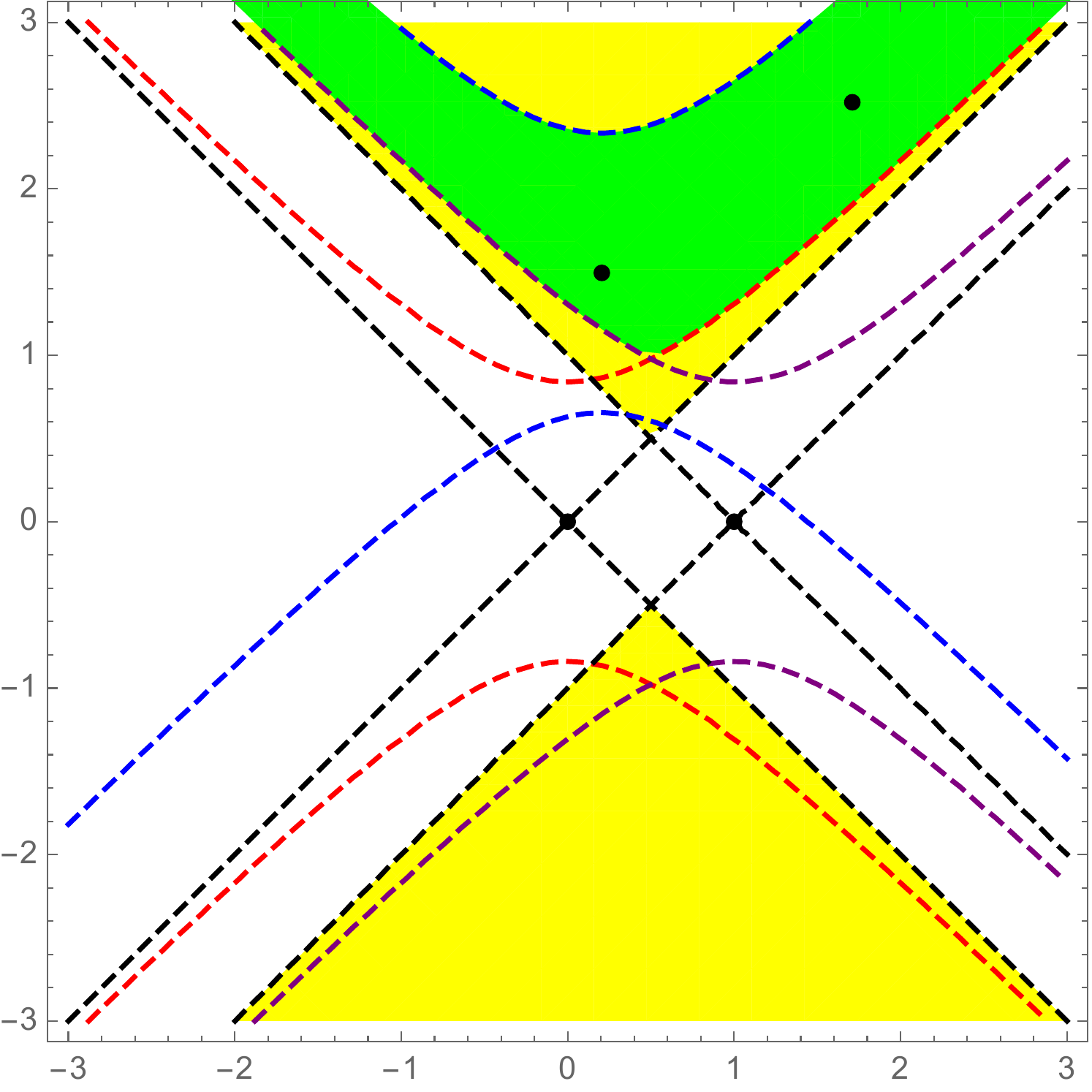}};
   \tiny  
   
   \node at (-.21,0.21){$y_1$};  
   \node at (1.15,0.21){$y_3$};
    \node at (0.2,2.2){$y_2$};
    \node at (2.27,3.63){$y_4$};
    \normalsize
\end{tikzpicture}
    \end{minipage}
    \caption{Configuration $\boldsymbol{w}_b$. $y_2$ can be placed anywhere in yellow region (which overlaps with green region), after which $y_4$ must be placed in green region to enforce $\text{sgn}(k(\boldsymbol{w}_b))=(++++-\hspace{0.12cm}-)$. The left picture corresponds to $h=0$ and the right picture to $h\neq 0$. For $h=0$ for this choice of $y_2$ placing $y_4$ in the left green region gives $z,\bar{z}\in(-\infty,0)$, and placing $y_4$ in right green region gives $z,\bar{z}\in(1,\infty)$. For $h\neq 0$ the green regions merge and $z,\bar{z}\in(-\infty,0), z,\bar{z}\in(0,1), z,\bar{z}\in(1,\infty)$ and $z,\bar{z}\in\mathbb{C}$ are all possible. The red, purple, and blue dashed lines correspond to the curves $y_{14}^2=0, y_{34}^2=0,$ and $y_{24}^2=0$ and reduce to light cones at $h=0$. $z,\bar{z}\in(0,1)$ is found to be `rare' and occurs only when $y_4$ is placed near the intersection of the red and purple curves.}
    \label{figg26}
\end{figure}
\newline
\newline
It is also worth commenting about the `kinematics reversed' case, where $\boldsymbol{w}\in \bar{V}_1$ and $\text{sgn}(k(\boldsymbol{w}))=(++++-\hspace{0.12cm}-)$. In this case $z,\bar{z}\in(-\infty,0), z,\bar{z}\in(0,1), z,\bar{z}\in(1,\infty)$ or $z,\bar{z}\in\mathbb{C}$ are all possible. Similarly to above, any such configuration can be conformally mapped to a simpler configuration $\boldsymbol{w}_b$ without changing the signs of the kinematics
\begin{align}
    \boldsymbol{w}_b=(z_1,z_2,z_3,z_4)=\left(0,\frac{1}{2}(v_2+u_2,0,0,v_2-u_2),e_3,\frac{1}{2}(v_4+u_4,0,h,v_4-u_4)\right).
\end{align}
When $h=0$ in this case it is again easily proved that only $z,\bar{z}\in (-\infty,0)$ and $z,\bar{z}\in (1,\infty)$ are possible (figure \ref{figg26} left). For $h\neq 0$ each of $z,\bar{z}\in(-\infty,0), z,\bar{z}\in(0,1), z,\bar{z}\in(1,\infty)$ or $z,\bar{z}\in\mathbb{C}$ are all possible (figure \ref{figg26} right). Configurations with $z,\bar{z}\in (0,1)$ are observed to be much rarer than the other cases.

\section{Permutations}\label{permutation}
\setlength{\extrarowheight}{1pt}
\begin{table}[h!]
\begin{center}
 \begin{tabular}{||c|c|c|c||} 
 \hline
 $\sigma_i\subset S_4$ & $\rho_i$ & $z$ transform &  $I$ transform\\ [0.5ex] 
 \hline

 $\sigma_1=\{(\sp\sp),(12)(34),(13)(24),(14)(23)\}$ & $\begin{pmatrix}1&0\\0&1 \end{pmatrix}$ & $z\rightarrow z$& $I\rightarrow I$ \\

 \hline
$\sigma_2=\{(24),(13),(1234),(1432)\}$ & $\begin{pmatrix}1&-1\\0&-1 \end{pmatrix}$ & $z\rightarrow 1-z$&  $I\rightarrow I$ \\

 \hline
 $\sigma_3=\{(23),(14),(1243),(1342)\}$ & $\begin{pmatrix}0&1\\1&0 \end{pmatrix}$ & $z\rightarrow \frac{1}{z}$& $I\rightarrow u\sp I$ \\

 \hline
 $\sigma_4=\{(12),(34),(1423),(1324)\}$ & $\begin{pmatrix}-1&0\\-1&1 \end{pmatrix}$ & $z\rightarrow \frac{z}{z-1}$& $I\rightarrow v\sp I$ \\

 \hline
 $\sigma_5=\{(234),(124),(132),(143)\}$ & $\begin{pmatrix}0&-1\\1&-1 \end{pmatrix}$ & $z\rightarrow \frac{1}{1-z}$& $I\rightarrow u\sp I$ \\

 \hline
 $\sigma_6=\{(243),(123),(134),(142)\}$ & $\begin{pmatrix}1&-1\\1&0 \end{pmatrix}$ & $z\rightarrow \frac{z-1}{z}$&  $I\rightarrow v\sp I$ \\

 \hline
\end{tabular}
\caption{Transformation of box integral and conformal invariants under permutations.}
\label{table5}
\end{center}
\end{table}\noindent Let $\boldsymbol{U}=(U_1,U_2,U_3,U_4)\in V$ be a configuration and $\si\boldsymbol{U}\equiv(U_{\si(1)},U_{\si(2)},U_{\si(3)},U_{\si(4)})$ for $\si\in S_4$ denote the permuted configuration. $S_4$ generates transformations of the conformal invariants $(z,\bar{z})\rightarrow(z',\bar{z}')=(f(z),f(\bar{z}))$ for the same function $f$. Note that since we take $\bar{z}\geq z$ we take $z'=f(\bar{z})$ and $\bar{z}'=f(z)$ if $f(z)>f(\bar{z})$. This $S_4$ action can be represented by a two-dimensional projective representation $\rho:S_4\rightarrow PSL(2,\mathbb{F})$, where $\mathbb{F}$ is the field associated to the conformal invariants $z$ and $\bar{z}$, defined by
\begin{align}
    \rho(\sigma)=\begin{pmatrix}a&b\\c&d\end{pmatrix},\spac \rho(\sigma)\begin{pmatrix}z\\1\end{pmatrix}=\begin{pmatrix}\frac{az+b}{cz+d}\\1\end{pmatrix}.
\end{align}
Note that $\rho$ is not one-to-one and so the representation is not faithful. $S_4$ splits into six distinct subsets $\sigma_i\subset S_4$ of order four, on which $\rho$ takes a single value $\rho_i\equiv\rho([\sigma_i])$, where $[\sigma_i]$ denotes a representative from $\sigma_i$ (table \ref{table5}). The kernel of this transformation is $\si_1\simeq V_4$, where $V_4$ is the Klein four-group. $V_4$ is a normal subgroup of $S_4$ and $\Im(\rho)\simeq S_4/V_4 \simeq S_3$. Therefore we can faithfully represent $S_4$ on the $z$ variables as $\rho_i$ for $i=1,2,\dots,6$, which we recognise as the two-dimensional irreducible representation of $S_3$. The action of $S_4$ on the conformal invariants $z$ and $\bar{z}$ is also known as the anharmonic group. In table \ref{table5} we also show how the integral (\ref{5b}) responds to permutations of the external points for finite configurations. It is not invariant but transforms by a conformally invariant factor $1$, $u$, or $v$.
\section{Double Infinity Calculation}\label{doubleinfcalcdetails}
The box integral (\ref{12i}) becomes $I_{\boldsymbol{X}}^{+-}(\xi_+,\xi_-)=$
\begin{align}
    \int_{x,r,t} \frac{\frac{2i}{\pi}r^2}{(t\!-\!r\!+\!i\epsilon)(t\!+\!r\!-\!i\epsilon)(t\!-\!1\!-\!r\!+\!i\epsilon)(t\!-\!1\!+\!r\!-\!i\epsilon)(t\!+\!rx\!-\!\xi_+\!+\!i\epsilon')(t\!-\!rx\!-\!\xi_-\!-\!i\epsilon')}.
\end{align}
There are six poles of the integrand in the $t$ plane, three in the upper half-plane and three in the lower half-plane. We take a semicircular contour, closed at infinity in the upper half-plane, so only the poles $t_1=-r+i\epsilon,\spa t_2=-r+1+i\epsilon,\spa t_3=rx+\xi_-+i\epsilon'$ contribute to the integral. Performing the $t$ integral with the residue theorem we recover
\begin{align}
     I_{\boldsymbol{X}}^{+-}(\xi_+,\xi_-)=\int_{x,r}\bigg(\frac{r}{(r-(-\frac{1}{2}+i\epsilon))(r-(\frac{-\xi_++i\epsilon}{1-x}))(r-(\frac{-\xi_-+i\epsilon}{1+x}))}\end{align}
    \begin{align*}
    -\frac{r}{(r-(\frac{1}{2}+i\epsilon))(r-(\frac{1-\xi_++i\epsilon}{1-x}))(r-(\frac{1-\xi_-+i\epsilon}{1+x}))}
    \end{align*}
    \begin{align*}
    -\frac{2r^2}{(r-(\frac{-\xi_-+i\epsilon}{1+x}))(r-(\frac{\xi_-+i\epsilon}{1-x}))(r-(\frac{1-\xi_-+i\epsilon}{1+x}))(r-(\frac{\xi_--1+i\epsilon}{1-x}))(r-(\frac{\Delta\xi-i\epsilon'}{2x}))}\bigg), 
\end{align*}
where $\int_{x,r}\equiv \int_{-1}^1dx\int_0^\infty dr$ and $\Delta\xi\equiv \xi_+-\xi_-$. To perform the $r$ integral we use a keyhole contour $C$, which comes in from $-\infty$ to 0 above the real axis, travels clockwise in an infinitesimal circle around $0$, and leaves from 0 to $-\infty$ below the real axis. If $f(r)$ is a function such that $\log(r)f(r)$ vanishes sufficiently quickly near the origin and at infinity, and $\log(z)$ is the complex logarithm with the branch cut chosen on the negative real axis, then
\begin{align}
    \int_C f(z)\log(z)dz=\int_{-\infty}^0(\log(r)+i\pi)f(r)dr+\int_0^{-\infty}(\log(r)-i\pi)f(r)dr=2\pi i\int_{-\infty}^{0}dr f(r).
\end{align}
Furthermore using the residue theorem we can deduce a formula for the integral to be
\begin{align}
    \int_{-\infty}^0 f(r)dr=\sum_{k}\Res_{z=z_k}(\log(z)f(z)),
\end{align}
where the sum is over all poles in the complex $z$ plane. There are a number of cancellations upon computing the residues. Indeed making the substitution $r'=-r$ the contributions from the two poles at $r'=\frac{\xi_--i\epsilon}{1+x}$ cancel, as well as those from the poles at $r'=\frac{\xi_--1-i\epsilon}{1+x}$. Therefore after performing the $r'$ integral the expression becomes reasonably compact
\begin{align}
   I_{\boldsymbol{X}}^{+-}(\xi_+,\xi_-)=2\int_{-1}^1dx\Bigg(-\frac{\log(\frac{1}{2})}{(x-(1-2\xi_++i\epsilon x))(x-(-1+2\xi_-+i\epsilon x))} 
\end{align}
\begin{align*}
    -\frac{\log(-\frac{1}{2}-i\epsilon)}{(x-(-1+2\xi_+-i\epsilon x))(x-(1-2\xi_--i\epsilon x))}
\end{align*}
\begin{align*}
    +\frac{(\Delta \xi)^2}{(\Box \xi)^2(\Box \xi-2)^2}\frac{4x\log(\frac{-\Delta\xi+i\epsilon'}{2x})}{(x^2-(\frac{-\Delta\xi+i\epsilon x}{\Box\xi})^2)(x^2-(\frac{-\Delta\xi+i\epsilon x}{\Box\xi-2})^2)}
\end{align*}
\begin{align*}
+\bigg(\frac{\xi_+}{\Box\xi}\frac{\log(\frac{\xi_+-i\epsilon}{1-x})}{(x-(1-2\xi_++i\epsilon x))(x-(\frac{-\Delta\xi+i\epsilon x}{\Box\xi}))}
\end{align*}
\begin{align*}
    +\frac{\xi_-}{\Box\xi}\frac{\log(\frac{-\xi_--i\epsilon}{1-x})}{(x-(1-2\xi_--i\epsilon x))(x-(\frac{\Delta\xi-i\epsilon x}{\Box\xi}))}+(\xi_\pm\rightarrow 1-\xi_\mp)\bigg)    \Bigg),
\end{align*}
where we defined $\Box\xi\equiv \xi_++\xi_-$. Note that this expression is checked to be invariant under $\xi_\pm\rightarrow 1-\xi_\mp$ after noting that under this replacement $\Delta\xi\rightarrow\Delta\xi$ and $\Box\xi \rightarrow 2-\Box\xi$. It would be nice to directly integrate the above expression into logs/dilogs, however there is still a small subtlety. The terms $i\epsilon x$ are not of definite sign over the integration range, which is not desirable as the $i\epsilon$ should specify the branch of our final expression. Therefore we split the $x$ integral from $-1$ to $0$ and from $0$ to $1$. As a toy example we would have
\begin{align}
    \int_{-1}^1\frac{dx}{x-(a+i\epsilon x)}=\int_{-1}^0\frac{dx}{x-(a-i\epsilon)}+\int_{0}^1\frac{dx}{x-(a+i\epsilon)}
\end{align}
\begin{align*}
    =\int_{0}^1dx\left(\frac{-1}{x-(-a+i\epsilon)}+\frac{1}{x-(a+i\epsilon)}\right).
\end{align*}
We perform this procedure to each of the five terms in the integral above. It is useful to introduce a shorthand for the combinations that appear in the denominators, so we define
\begin{align}\label{u1}
  r_{1\pm}\equiv 1-2\xi_+\pm i\epsilon,\spac r_{2\pm}\equiv-1+2\xi_-\pm i\epsilon,
\end{align}
\begin{align}\label{u2}
    s_{1\pm}\equiv\frac{-\Delta\xi\pm i\epsilon}{\Box\xi},\spac s_{2\pm}\equiv\frac{-\Delta\xi\pm i \epsilon}{\Box\xi-2}.
\end{align}
Furthermore for the third term we change variables $x^2=y$ and split up the logarithm. The resulting expression for the box integral is
\begin{align*}
    I_{\boldsymbol{X}}^{+-}(\xi_+,\xi_-)=2\int_0^1dx\Bigg(-\frac{\log(\frac{1}{2})}{(x-r_{1+})(x-r_{2+})}-\frac{\log(\frac{1}{2})}{(x+r_{1-})(x+r_{2-})}-\frac{\log(-\frac{1}{2}-i\epsilon)}{(x+r_{1+})(x+r_{2+})}
\end{align*}
\begin{align*}
    +\frac{(\Delta \xi)^2}{(\Box \xi)^2(\Box \xi-2)^2}\bigg(\frac{2\log(-\frac{\Delta\xi}{2}+i\epsilon')-\log(x)}{(x-s^2_{1+})(x-s^2_{2+})}+\frac{-2\log(\frac{\Delta\xi}{2}-i\epsilon')+\log(x)}{(x-s^2_{1-})(x-s^2_{2-})}\bigg)
\end{align*}
\begin{align*}
   -\frac{\log(-\frac{1}{2}-i\epsilon)}{(x-r_{1-})(x-r_{2-})} +\bigg(\frac{\xi_+}{\Box\xi}\bigg\{\frac{\log(\xi_+-i\epsilon)-\log(1-x)}{(x-r_{1+})(x-s_{1+})}+\frac{\log(\xi_+-i\epsilon)-\log(1+x)}{(x+r_{1-})(x+s_{1-})}\bigg\}
    \end{align*}
    \begin{align}
    +\frac{\xi_-}{\Box\xi}\bigg\{\frac{\log(-\xi_--i\epsilon)-\log(1-x)}{(x+r_{2+})(x+s_{1+})}+\frac{\log(-\xi_--i\epsilon)-\log(1+x)}{(x-r_{2-})(x-s_{1-})}\bigg\}+(\xi_\pm\rightarrow 1-\xi_{\mp})\bigg)\Bigg).
\end{align}
The remaining integrals may now be safely computed in terms of logs and dilogs, and the final answer is given in (\ref{7i}). The calculation for the configuration $\boldsymbol{X}^{--}$ is easier, because if we perform the $t$ integral by closing the contour in the upper half-plane, there are only two poles to consider. The final answer is given in (\ref{8i}). 

\addtocontents{toc}{\protect\setcounter{tocdepth}{-1}}
\printbibliography[heading=bibintoc,title={References}]

\end{document}